\pdfoutput=1

\documentclass{iopart}
\usepackage{graphicx} 
\begin{document}

\topical{Origin of two-dimensional electron gases at oxide interfaces: insights from theory}

\author{N C Bristowe$^{1,2}$, Philippe Ghosez$^{1}$, P B Littlewood$^{2,3,4}$ 
and Emilio Artacho$^{2,5,6}$}

\address{$^1$ Theoretical Materials Physics, University of Li\`ege,
             B-4000 Sart-Tilman, Belgium}

\address{$^2$ Theory of Condensed Matter,
             Cavendish Laboratory, University of Cambridge, 
             J. J. Thomson Avenue, Cambridge CB3 0HE, UK}

\address{$^3$ Physical Sciences and Engineering,
             Argonne National Laboratory, 
             Argonne, Illinois 60439, USA} 
 
\address{$^4$ James Franck Institute, University of Chicago,
             929 E 57th Street, Chicago IL 60637, USA} 
             
\address{$^5$ CIC Nanogune, and DIPC,
		Tolosa Hiribidea 76, 20018 San Sebastian, Spain}  
	
\address{$^6$ Basque Foundation for Science Ikerbasque, 48011 Bilbao, Spain}

\ead{n.bristowe@ulg.ac.be}
\begin{abstract}
  The response of oxide thin films to polar discontinuities at interfaces
and surfaces has generated an enormous activity due to the variety of
interesting effects it gives rise to.
  A case in point is the discovery of the electron gas at the interface
between LaAlO$_3$ and SrTiO$_3$, which has since been shown to 
be quasi-two-dimensional, switchable, magnetic and/or 
superconducting. 
  Despite these findings, the origin of the two-dimensional electron 
gas is highly debated and several possible mechanisms remain. 
  Here we review the main proposed mechanisms and attempt to 
model expected effects in a quantitative way with the ambition of 
better constraining what effects can/cannot explain the observed 
phenomenology.
  We do it in the framework of a phenomenological model for 
understanding electronic and/or redox screening of the chemical 
charge in oxide heterostructures.
  We also discuss the effect of intermixing, both conserving and 
non-conserving the total stoichiometry.

\end{abstract}




\section{Introduction}

  Oxide interfaces have received considerable attention within 
the past couple decades.
  From the device perspective, many technological applications 
depend crucially on interfacial properties and the fundamental 
understanding at the microscopic level is often critical for optimising 
device performance. 
  In particular oxide interfaces have also offered the opportunity 
to discover and study completely novel, and often unexpected, 
fundamental materials physics. 
  The LaAlO$_3$-SrTiO$_3$ (LAO-STO) interface has become a 
prototypical example of such a system.
  In bulk form both oxides are electronically fairly simple 
non-magnetic insulators.
 Remarkably, the interface between 
the two can exhibit metallic~\cite{Ohtomo2004}, 
magnetic~\cite{Brinkman2007} and/or 
superconducting~\cite{Reyren2007,Bert2011,Li2011} 
behaviour of a quasi-two-dimensional character~\cite{Basletic2008}.
  It has since been proposed that the LAO-STO system may find 
applications in field effect devices~\cite{Thiel2006,Cen2008}, 
sensors~\cite{Xie2011}, nano-photodetectors~\cite{Irvin2010}, 
thermoelectrics~\cite{Pallecchi2010,Filippetti2012},
and solar cells~\cite{Assmann2013,Liang2013}.
  The combination of exotic physical properties and wide-range of 
potential device applications has made the LAO-STO interface a 
highly popular system of study with several dedicated 
reviews and perspectives~\cite{Hwang2006,Pauli2008,
Huijben2009,Mannhart2010,Pentcheva2010,Chen2010,
Schlom2011,Chambers2011,Pentcheva2012,Gabay2013}.

  Despite the enormous amount of activity on this system, 
the origin of a two-dimensional electron gas (2DEG) at the 
LAO-STO interface remains unclear. 
  Arguably the most popular explanation, called the 
`electronic reconstruction', is based on the concept of a 
`polar catastrophe'~\cite{Nakagawa2006}.
 The concept uses basic electrostatics 
(see section~\ref{electrostatics}) to show that a net charge at a polar, insulating 
interface such as LAO-STO produces a finite non-decaying 
electric field.
  This diverging electrostatic potential is clearly 
energetically unstable upon increasing thickness of the oxide film.
  In reality the system requires some form of charge compensation.
  This might occur through an electronic reconstruction (see 
section~\ref{Ereconstruction}) whereby the electric field produces 
an electronic transfer similar to Zener tunnelling 
from the surface valence band to the interface conduction band. 
  This  process should not be taken too literally.
  The kinetic process of establishing the 2DEG can be more
involved and is beyond the scope of this paper. 
  The polar catastrophe and electronic reconstruction concepts
establish the paradigm in an intuitive way.
  It is interesting, however, that the same mechanism can be seen
from a different perspective~\cite{Janotti2012}.
 There it was noted that the isolated neutral interface and surface should have free carriers 
associated to them to start with, which (partly) annihilate when 
brought close to each other.
  It is an alternative view of the same mechanism, with the same
predictions for the equilibrium result, but different perspective on the
process.
 
  Alternative explanations suggest that charge compensation 
may be achieved through chemical modifications or chemical 
plus electronic (redox) modifications (see section~\ref{sec: redox}).
  A third category of proposed mechanisms is not based on polarity 
arguments but instead extrinsic effects induced by the film growth 
process such as doping through oxygen vacancies or interface 
cation intermixing (see section~\ref{sec: interface}).
  Clearly the origin of the 2DEG is not just of academic interest, 
but key to understand what other systems might show similar 
behaviour and how one might be able to tune device properties.
 
  A large number of experimental and theoretical studies 
have been devoted to clarify which mechanism(s) 
operates in LAO-STO.   
  A recent short review attempted to summarise the results, 
concluding that the electronic reconstruction mechanism 
appears to support experiment~\cite{Schlom2011}. 
  The electronic reconstruction mechanism is certainly 
strongly supported by several experimental observations, 
most notably the observation of a critical thickness of the 
LAO film for the onset of interface conductivity~\cite{Thiel2006} 
which can be tuned~\cite{Reinle-Schmitt2012}.
 However the electronic reconstruction mechanism does 
not naturally explain all experimental observations. 
  For example the discovery of a sizeable density of (trapped) 
Ti 3$d$ like states below the critical thickness~\cite{Sing2009,
Berner2010,Fujimori2010,Rubano2011,Slooten2013}, the lack of surface 
hole carriers~\cite{Thiel2006} or states near the Fermi 
level~\cite{Berner2013a,Plumb2013}, the almost negligible 
electric field within LAO below the critical 
thickness~\cite{Segal2009a,Slooten2013,Berner2013b}, and the 
apparent disappearance of conductivity at any LAO 
thickness for samples grown at high Oxygen partial 
pressures~\cite{Herranz2007,Kalabukhov2011}.
  It is apparent that the situation is not so clear. 
   
  The main aim here is to review the findings for this and 
related systems in the framework of a simple model.
 This model describes simply 
but quantitatively the several effects that concur in the 
physics of such systems. 
  In doing so, we hope to better constrain 
what effects can/cannot explain the observed phenomenology.
  It is therefore limited in scope and far from any ambition of being
exhaustive in the review of the activity around these systems.
  In section~\ref{polar interfaces} we briefly introduce the 
concept of polar interfaces, highlighting for the case of LAO-STO 
the basic theory behind determining the interface net charge 
and resulting electrostatics. 
  We then begin by exploring the various proposed mechanisms 
behind the origin of 2DEGs at oxide interfaces. 
  We present a phenomenological model for understanding electronic 
and/or redox screening of the chemical charge in polar 
heterostructures in sections~\ref{Ereconstruction} and~\ref{sec: redox}.
  In section~\ref{sec: interface} we discuss the effect of intermixing, 
both conserving and non-conserving the total stoichiometry.




\section{Polar interfaces \label{polar interfaces}}

\subsection{The net charge \label{net charge}}

\begin{figure}[t]
\begin{center}
\includegraphics[width=0.9\textwidth]{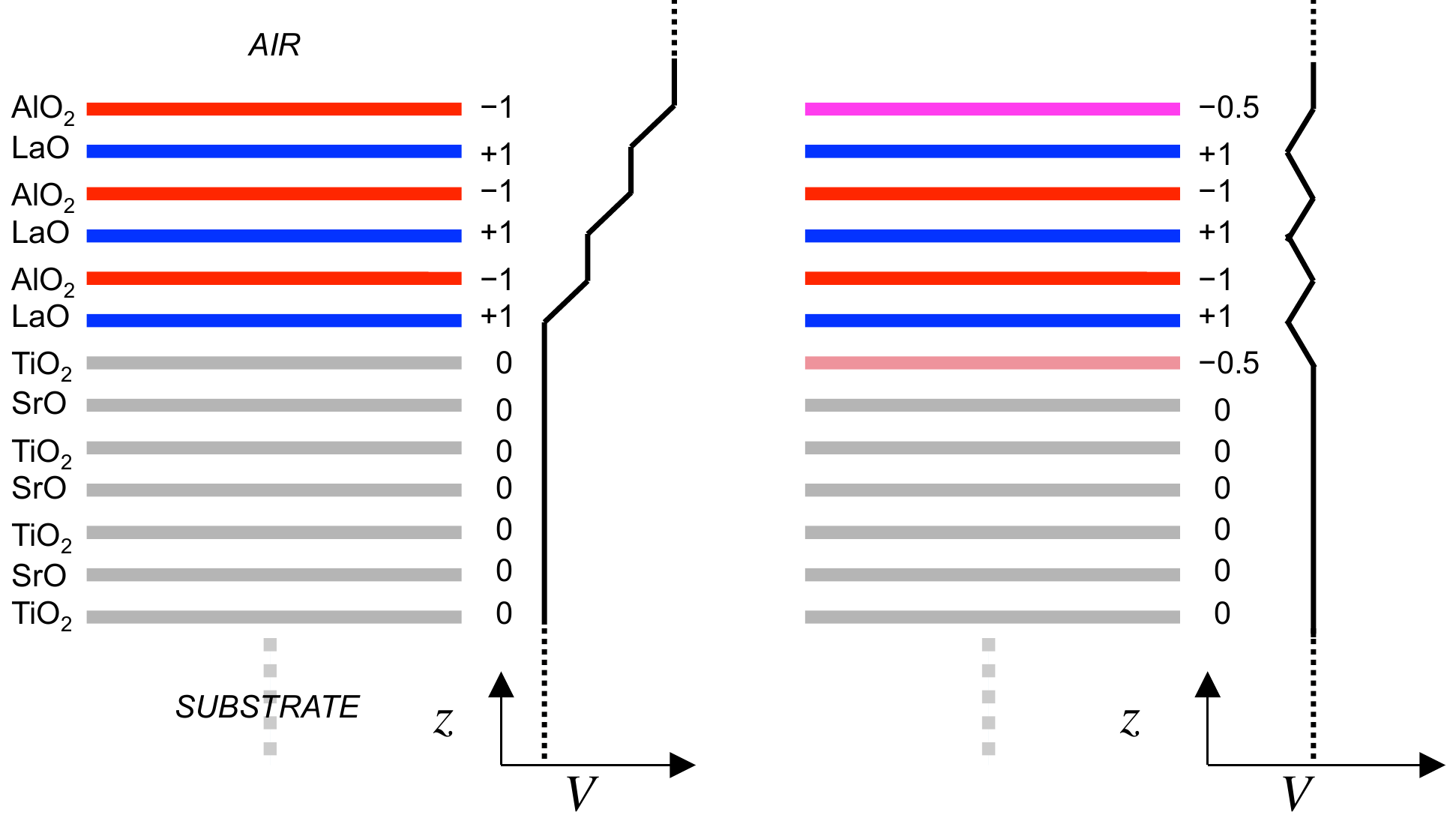}
\caption{\label{SKETCH}{Electrostatic model illustrating  the
polar catastrophe based on atomic layers taken as 
charged planes with the net charge (in units of $e/\Theta$) 
given by ionic formal charges. 
The $z$ axis lies along the [001] 
crytallographic direction, which is normal to the surface. 
The left panel depicts a pristine interface, together with 
the electrostatic potential, $V$,
it generates, which tends to diverge with increasing LAO
film thickness. The right panel shows the stabilised system
for which $-0.5 e/\Theta$ has been transferred from the
surface layer to the interface.
 }}
\end{center}
\end{figure}

  In the seminal paper of Ohtomo and Hwang~\cite{Ohtomo2004} 
the discovery of the two-dimensional electron gas at the interface 
between the two nominally insulating oxides, LAO and STO, 
was rationalised in terms of the electrostatics of the formal 
charges on each ion, i.e. -2$e$ for O, +4$e$ for Ti and so on,
as illustrated in Fig.~\ref{SKETCH}.
  From these formal charges, the (001) layers in STO (SrO and 
TiO$_2$) are charge neutral whilst in LAO (LaO and AlO$_2$) 
they are +1$e$ and -1$e$ per layer formula unit. 
  At the (001) LAO-STO interface, taking the two unperturbed 
bulk materials~\footnote{The polarisation response and electronic 
and redox screening of polar interfaces will be discussed in 
sections~\ref{electrostatics},~\ref{Ereconstruction} 
and~\ref{sec: redox}, respectively.}, these layer charges give 
rise to an imbalance at the interface of $\pm$ 0.5 $e/\Theta$. 
  The sign depends on the interface termination, and $\Theta$ 
is the 2D interface unit cell area corresponding to one formula unit. 
  These basic arguments, built upon the formal ionic charges, 
are at the heart of almost all subsequent work on this system.  

  It is tempting to suggest a renormalisation of these 
numbers by introducing covalency factors in an attempt 
to correct for realistic bonding in these far from ideally 
ionic $AB$O$_3$ materials. 
  It is well known that in most perovskites there is a sizeable 
covalent bonding character, especially in the $B$-O bonds. 
  One would then find after repeating the basic procedure 
above an arbitrary value of the interface charge smaller 
than $\pm$ 0.5 $e/\Theta$. 

  The first point we wish to highlight here is that when 
determining the net interface charge in these materials 
it would be incorrect to include covalency considerations 
such as that just mentioned (see ref.~\cite{Bristowe2011} 
for further discussion).
  The net interface charge of these band insulators is exactly 
as obtained from the simple counting of the formal charges 
on each species irrespective of covalency~\footnote{In crystals 
lacking a symmetry forbidding local dipoles on the atoms,
the formal ionic charges are not sufficient for determining 
the surface/interface charge~\cite{Stengel2011c}. In 
such cases the polarisation analysis explained in section~\ref{polarisation}
using the modern theory of polarisation would be correct, 
while an analysis based on formal charges alone 
would not.}. 
  For LAO-STO, the net interface charge is precisely 
0.5 $e/\Theta$~\cite{Bristowe2009,Stengel2009}. 

  Let us discuss this further. 
  We outline the arguments proposed in 
refs.~\cite{Stengel2009, Vanderbilt1993,Bristowe2009,
Bristowe2011,Stengel2011c}. 
  For electrostatics purposes (see next section) the calculation 
of the resulting electric field from an interface requires the 
knowledge of both the area density of free charges and 
the total polarisation, or bound charges (including the 
electric field response - see next section) of each material.
  The subtlety of the problem comes from what is meant by 
free charges and the definition of the polarisation.
  Remember we are considering the unperturbed bulk 
interface, so zero free charges in a strict sense. 
  However the free charge can also be defined as anything 
unaccounted for by the polarisation, and previously called 
chemical~\cite{Bristowe2009} or compositional 
charge~\cite{Murray2009a}. 
  The key to the problem is in the definition of the reference 
unperturbed (zero of) polarisation: altering its value will 
change the free charge by the same amount. 
  In the end it is the sum of the two that matters and is 
unaffected. 
  Centrosymmetric materials such as LAO and STO allow 
for a natural definition of zero polarisation, the {\it effective} 
polarisation then being the change of polarisation after some 
perturbation from the centrosymmetric reference. 
  Taking the polarisation of STO and LAO as zero, one can 
then use a charge counting technique such as the 
'polarisation-free' unit cell method~\cite{Goniakowski2008} 
to find a free (or chemical/compositional) interface charge 
density of 0.5 $e/\Theta$. 

\subsection{Polarisation and quanta \label{polarisation} }

  An alternative viewpoint, that is perhaps more elegant 
and rigorous, has been proposed~\cite{Stengel2009} 
which considers the {\it formal} polarisation instead of 
the {\it effective} value above.
  The formal polarisation is obtained by direct application 
of the modern theory of polarisation~\cite{Kingsmith1993}, 
for example by taking the ionic cores and charge centre of 
Wannier functions~\cite{Marzari1997} as point charges.
  The difference between the two definitions, effective and formal,
for the case of LAO-STO is exactly 0.5 $e/\Theta$: whilst the 
effective value for LAO and STO is zero, the formal values are 
0.5 $e/\Theta$ and zero, respectively.
  Within this definition the free charges are now zero, 
but the sum of free and bound charges remains 
0.5 $e/\Theta$ and hence the resulting electrostatic 
analysis is unaltered.   

  We have neglected to mention so far one more subtlety. 
  In a system of discrete point charges the bulk polarisation is 
defined only up to quanta of polarisation.
  When the definition of the bulk unit cell is displaced along 
a crystallographic direction and reaches one of the point 
charges, it disappears there and reappears at the other side 
of the unit cell.
  At this point the obtained polarisation jumps by a finite 
value of 
\begin{equation*}
P_0 = 1 \, e L / V = 1 \, e/\Theta,
\end{equation*}
which is called a quantum of polarisation,
where $L$ is the length of the cell along the direction
of the flip, and $V$ is the cell volume (polarisation as
dipole per unit volume).
  Clearly there is no \textit{a} \textit{priori} best 
choice of the unit cell as long as the surface is unknown. 
  The bulk polarisation value is therefore only defined modulo 
polarisation quanta.
  In centrosymmetric perovskites two sets of formal 
polarisation values are allowed, 
\begin{equation*}
\{...,-2P_0,-P_0,0,P_0,2P_0,...\}
\end{equation*}
as in STO, and 
\begin{equation*}
\{...,-3P_0/2,-P_0/2,P_0/2,3P_0/2,...\}
\end{equation*}
as in LAO, 
since these are the only two sets that are invariant 
under change of sign. 
  This is the key difference with the effective polarisation 
as defined above, where all centrosymmetric materials 
were defined to have 0 polarisation: now they can have 
0 or $P_0/2$, both modulo $P_0$. 

\subsection{Symmetry and topology \label{topology} }

\begin{figure}[t]�
\begin{center}
\includegraphics[width=0.7\textwidth]{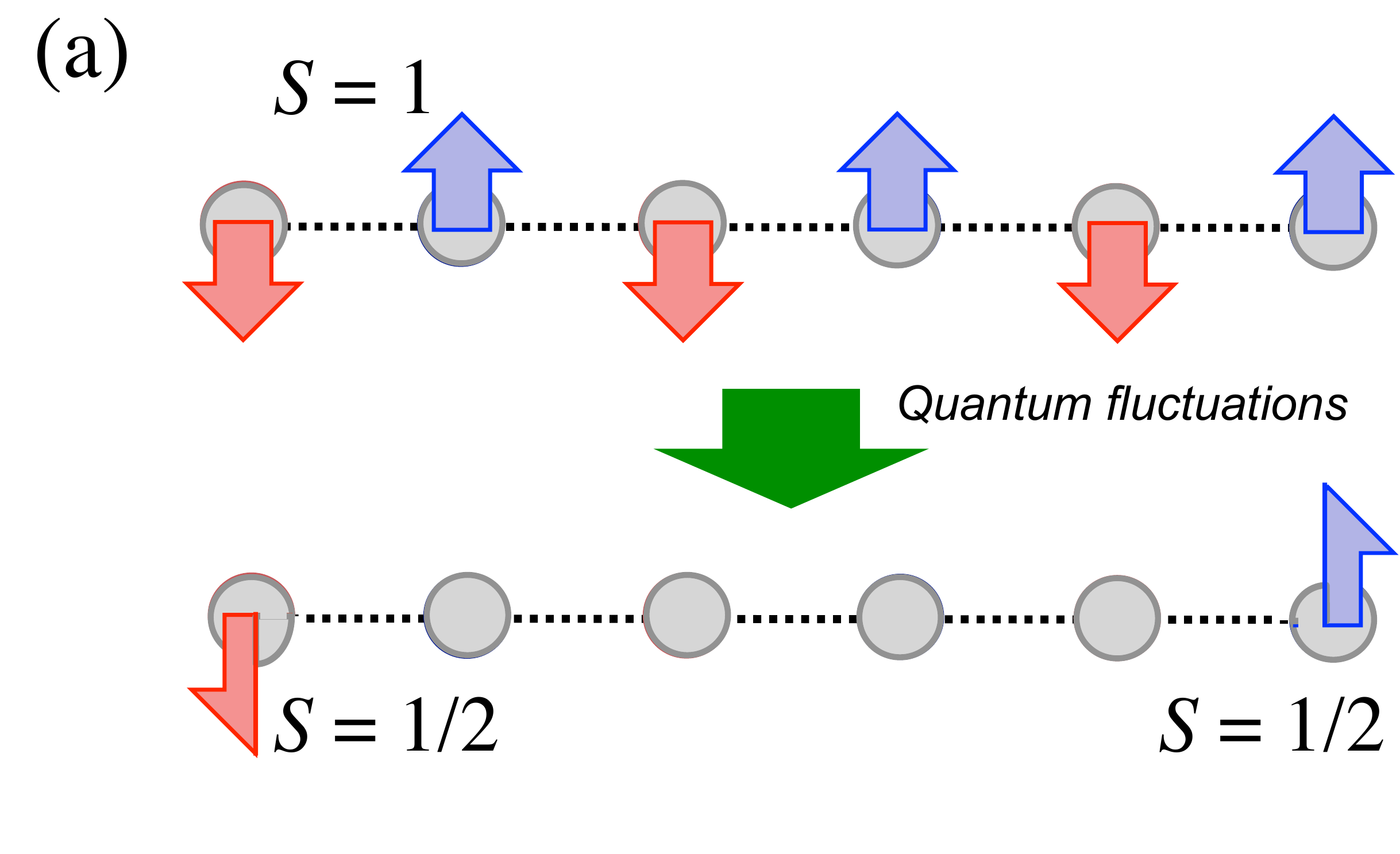}
\includegraphics[width=0.7\textwidth]{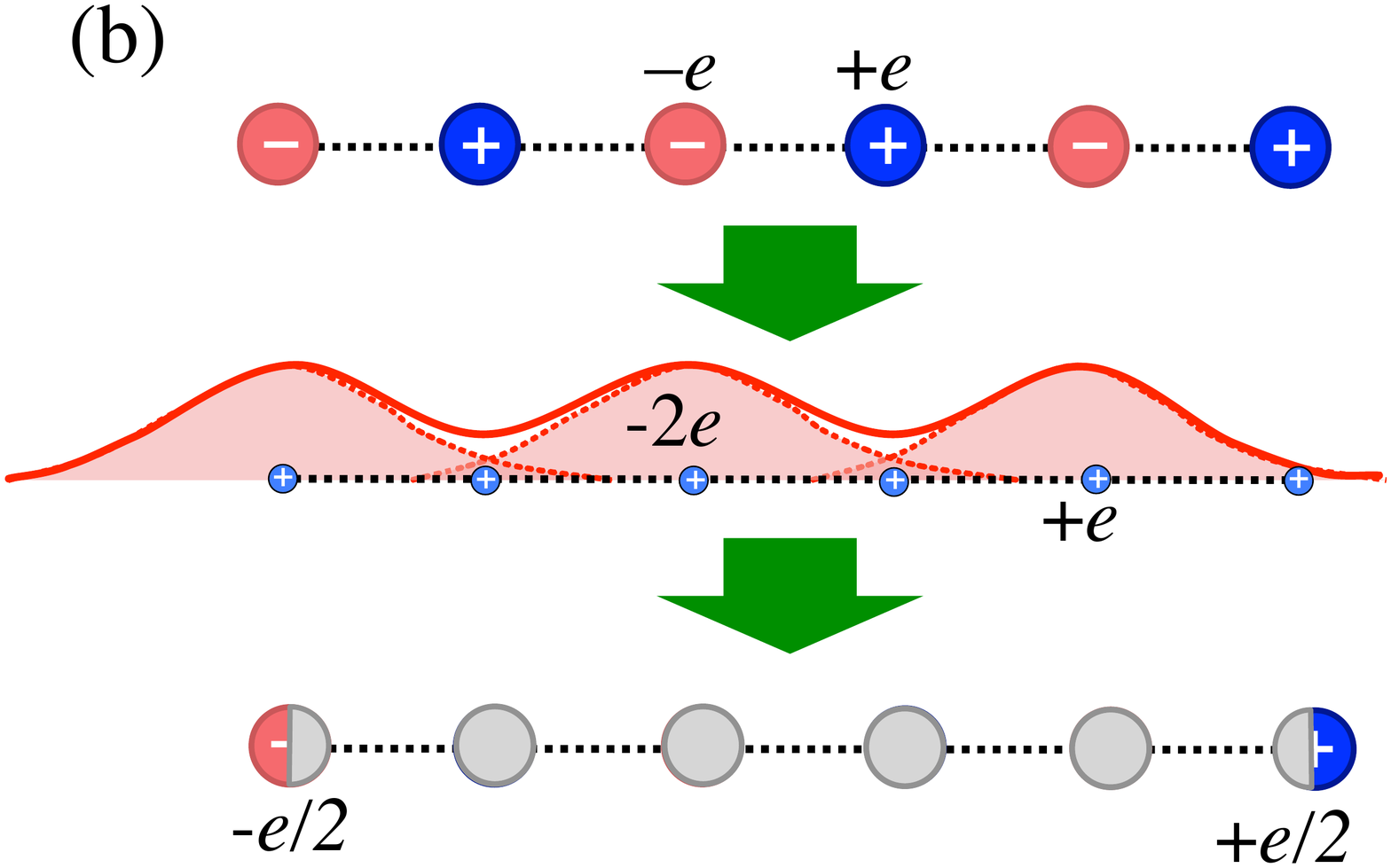}
\caption{\label{CHAIN}{Symmetry protected topological 
phases: (a) Sketch of a finite Haldane chain, a quantum 
antiferromagnet of Heisenberg spins with $S=1$, which
displays net $S=1/2$ at the ends. It is adapted from the
figure in Ref.~\cite{Qi2012}. (b) Analogous sketch for
a chain of quantum charges, taking delocalised electrons
and localised nuclei. The middle panel shows the quantum
delocalisation of the former, and the corresponding Wannier
functions. The net effect is that of half the quantum charge
at each end.}}
\end{center}
\end{figure}

  We have approached the problem of determining the
polarisation from a perspective based on point charges 
using the centre of charge of the valence 
band Wannier functions~\cite{Marzari1997}.
  An alternative approach defines the bulk polarisation 
of an insulator in terms of a Berry 
phase~\cite{Kingsmith1993,Resta1994}.
  This formulation, called the modern theory of polarisation,
arises from the consideration
of the current that flows through the system when
adiabatically changing its polarisation state.
  In 1984 it was established by Berry~\cite{Berry1984}
that such an adiabatic change of a Hamiltonian
is associated with the appearance of a complex
phase in the wave-function that describes the system
as it adiabatically evolves. 
  This phase is of deep physical significance and
is called a geometric phase, which, in principle
can take any value, corresponding to an arbitrary
value of the polarisation.
  The presence of a symmetry pins the
polarisation (and corresponding phase) to particular
values, and the phase is referred to as topological.

  In centrosymmetric insulators like LAO and 
STO, the two possible values of the polarisation
0 and $P_0/2$ (both modulo $P_0$) described 
in section~\ref{polarisation} correspond
to the two values of the Berry phase, 0 and $\pi$, 
modulo $2\pi$.
  It is interesting to make the analogy with other 
symmetry-protected topological phases (in e.g.
topological insulators).
 Here the relevant symmetry is 
e.g. time-reversal, and the topological phase relates to
other properties, e.g. to magnetisation or its susceptibility.
  In the conventional table for topological insulators,
one axis corresponds to the relevant symmetry, 
its entries conventionally being time reversal, 
particle-hole and their product.
  Arguably, in this table categorising the different 
symmetry-protected topological phases,  spatial 
symmetries should also feature, 
which would then include the insulators considered 
in this review. 
  It has already been done for two-dimensional 
systems~\cite{Bristowe2012b,Jadaun2012}, referring to 
the quantisation of electric polarisation, as discussed here.
  Interestingly, for other symmetry groups, as for honeycomb
planar insulators (e.g. BN), there are more than two
sets of consistent values, three in this case~\cite{Bristowe2012b}.
  (There are further topological classifications of 
insulators based on crystal symmetries, but in a
different context to the one of this review~\cite{Fu2011}.)
  In analogy with other topological insulators,
the interface between two insulators of different
topological character in this sense, as LAO and STO,
gives rise to a 2DEG.
  Of course, such 2DEGs
are not topologically protected against any kind
of defect breaking the symmetry defining the phase.
This is analogous to the situation relating to a $Z_2$ topological insulator:
the 2DEG arising on its surface is not protected against
defects breaking the time-reversal symmetry, such
as magnetic impurities. 
  In this case, the 2DEG is affected by regular
disorder and defects.

  The analogy with other symmetry protected 
topological phases is best illustrated referring to a 
``perspective article" of X. Qi in Science in 
2012~\cite{Qi2012}.
 Here the concept of a symmetry-protected topological phase
is transmitted in terms of the Haldane chain, a one-dimensional 
Heisenberg model of localised quantum spins 
with S=1, interacting antiferromagnetically.
  In the ground state for a finite chain, the quantum
fluctuations of the local spins act in such a way that
the chain appears as if without spins, except for
a net spin of $S=1/2$ at one end and $S=-1/2$ 
at the other, in spite of their being no $S=1/2$
particles anywhere in the system.

  That system is beautifully analogous to a 
chain of alternating charges, $+$ and $-$ some
quantum of charge, which is again analogous
to the (100) atomic planes of LAO (see Figure 1
of ref.~\cite{Bristowe2011}).
  The net effect is that of half a quantum of 
charge of either sign at either end, as sketched in 
Fig.~\ref{CHAIN} (the quantum
fluctuations in this case manifest themselves on the
delocalisation illustrated in the figure for the electrons).

\subsection{Other interface planes \label{interface planes}}

  The discussion above was centred around the 
heterostructures grown in (001) planes because
most experiments have been performed for those
systems, in addition to the simplicity they offer.
  The generalisation to other growth directions
is simple, both using formal charges or using
polarisation.
  Using formal charges, the argument goes parallel
to the original one by Ohtomo and Hwang~\cite{Ohtomo2004}.
  The  stacking of [001] atomic layers 
in the ABO$_3$ perovskite is AO/BO$_2$/AO/BO$_2$,
and thus with formal charges $+1$, $-1$, $+1$, $-1$,
for LAO, while they are all neutral for STO.
  Similarly the stacking of [111] layers is
AO$_3$/B/AO$_3$/B, and therefore with alternating
$-3$ and $+3$ charges in LAO, and $-4$ and $+4$
for STO. 
  For [011] the stacking is ABO/O$_2$/ABO/O$_2$,
and the resulting formal charges are $+4$ and $-4$
for both materials. 
  One would thus expect a polarisation discontinuity
in [001] and [111] but not in [011].

\begin{figure}[t]
\begin{center}
\includegraphics[width=0.5\textwidth]{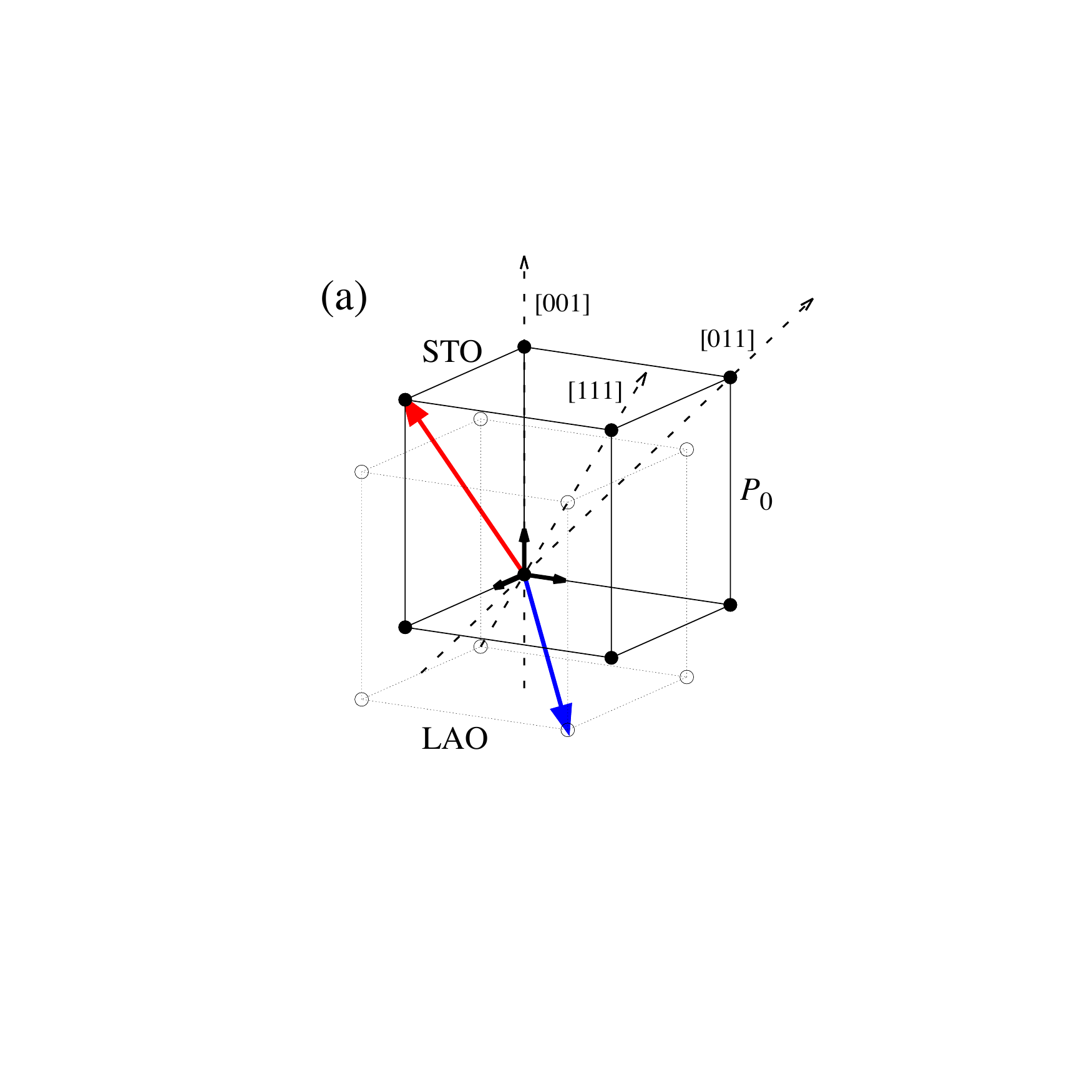}
\includegraphics[width=0.5\textwidth]{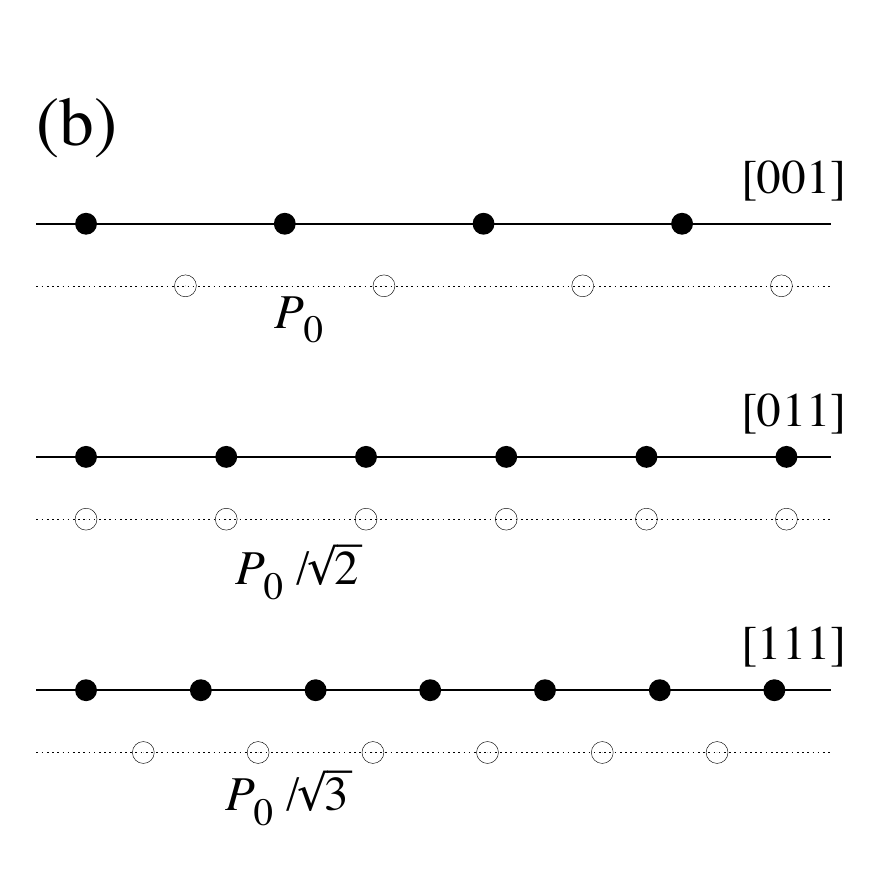}
\caption{\label{CUBES}{
  (a) Polarisation lattices for LAO and STO. 
  The circles represent the tips of the set (lattice) of 
polarisation vectors for LAO (empty circles) and 
for STO (filled). The origin of the vectors is at (0,0,0),
illustrated by a $\vec P$ vector for STO (red, lighter), 
and another for LAO (blue, darker).
  The cubic forms of LAO and STO have been assumed 
for simplicity, and the two lattice parameters ($P_0$, quantum
of polarisation) have been 
depicted as equal, although in reality $P_0^{LAO}$ is slightly 
different from $P_0^{STO}$.
  The LAO lattice of polarisation vectors does not include the 
origin ($\vec P = 0$), which lies in the centre of a cube, while 
the $\vec P$ lattice for STO  contains $\vec P = 0$.
  Dashed lines represent the [001], [011] and [111] 
directions, along which the lattice points are to be 
projected to see the polarisation values relevant for
the corresponding (001), (011) and (111) interfaces
or surfaces (i.e. the component perpendicular to the interface
plane).
  (b) Polarisation values obtained for the three directions
after the projection onto the corresponding axis of all points
found in the plane normal to the axis.
}}
\end{center}
\end{figure}

  Going back to the polarisation discussion, 
the arguments above generalise to 3D from the
1D set proposed in such a way that for 
one meaningful value of the polarisation there is 
a whole 3D lattice of $\vec P$ values. 
  In the STO case, and assuming the cubic phase for
simplicity, the lattice is simple cubic, with $P_0$ as 
lattice parameter, and such that $\vec P=0$ is
a member of the set. 
  For LAO the lattice is also simple cubic, but now the 
origin $\vec P=0$ is at the centre of a cube, and thus not
part of the set.
  This is illustrated in Fig.~\ref{CUBES} (a), and it
is analogous to the polarisation lattices already 
presented in the pioneering paper of Vanderbilt and
King-Smith~\cite{Vanderbilt1993}, and to the 
recent proposal for 2D honeycomb (graphenic) 
insulators in ref.~\cite{Bristowe2012b}.

  In order to predict for possible polarisation discontinuities
of these two materials in interfaces of arbitrary orientation,
one selects the direction normal to the interface 
and projects all $\vec P$ vectors of the lattice for 
each material onto the particular direction line.
  This is illustrated in Fig.~\ref{CUBES}, using the
[001], [011] and [111] directions as examples.
  The projections proposed in Fig.~\ref{CUBES} (a)
give rise to the 1D sets for the three directions, 
shown in Fig.~\ref{CUBES} (b), for both the STO
(continuum lines and filled circles) and the LAO
case (dotted lines and open circles).
  It is very apparent that both for (001) and (111) 
interfaces there are no possible choices of $\vec P$ 
for both materials such that $\Delta \vec P = 0$; 
there is always a discontinuity in the polarisation. 

  Consistently with this discussion and in addition
to the well-known 2DEG in the (001) case, a 2DEG 
has been recently discovered at the (111) LAO-STO 
interface~\cite{Herranz2012}.
  In this case, the atomic layers are alternating
$A$O$_3$ planes and $B$ planes, which for
LAO means -3 and +3 layers, while in STO
it means -4 and +4 layers. 
   There is therefore a discontinuity at the interface
corresponding to 0.5 $e/\Theta_{111}$.
   Noting that $\Theta_{111}=\sqrt{3} \Theta$ (for 
the $\Theta$ as defined in section~\ref{net charge}, for the (001) case)
we see that the formal charges argument, giving
a chemical charge of 0.5 $e/(\sqrt{3} \, \Theta)$, agrees
with what obtained from Figure~\ref{CUBES}, namely,
$\Delta P = 0.5 \, P_0 / \sqrt{3}$.

  While a discontinuity in the polarisation exists for 
(001) and (111) interfaces, it is not the case for the (011)
interface for which interfaces with no $\vec P$ 
discontinuity can be built. 
   This agrees with the consideration from formal charges
for which (011) planes of LAO (STO) are made
of LaAlO (SrTiO) planes alternating with O$_2$
planes, i.e. alternating planes of $+4$ $e/\Theta$ 
and $-4$ $e/\Theta$, in {\it both} cases.
  Therefore, as long as a O$_2$ plane always
separates $AB$O planes, there should be
no polarisation discontinuity.


  It is interesting, however, that experiments for 
(011) interfaces~\cite{Herranz2012,Annadi2013} 
show interface conduction 
similar to what found in the (001) case, in spite of the fact that
according to the discussion in the previous paragraph there should be
no polar discontinuity at the pristine interface, and thus
no intrinsic tendency to accumulate free carriers at
the interface.
  The same study~\cite{Annadi2013} arrives at an 
explanation of the
effect by a careful structural study of the interface,
which shows that it is not an ideal epitaxial (011)
interface, but rather a heavily stepped one, in a 
jigsaw pattern of (001) steps and terraces.
  These do present the polar discontinuity discussed
above, thereby explaining the appearance of the
2DEG.
  This raises the next question, as to why does 
this pattern emerge at all if the ideal interface would
represent no electrostatic penalty whatsoever.
  The answer as suggested in refs.~\cite{Bottin2005,Annadi2013}
is likely to be related to the fact that
although the ready-made \textit{interface} would not be
polar, the preformed ideal (011) \textit{surface} of STO 
is highly so and requires pre-screening, such as the observed
formation of the (001) steps and terraces,
before the growth of the film.


  Finally, the vector lattice above can also be used to
predict the polar discontinuities at stepped or vicinal 
surfaces, as demonstrated in \cite{Bristowe2012a}.
  The direction normal to the average interface 
defines the global polar discontinuity, although if
the interface is constituted by well-defined steps
separated by wide terraces, the contribution of 
steps and terraces can be distinguished, allowing
for the preparation of one-dimensional electron
gases along the steps~\cite{Bristowe2012a}.

\subsection{Removing the arbitrariness}

  In as much as we are defining the polar discontinuity
at an interface from the bulk polarisation values,
there is the indetermination in the lattice of quanta
described above. 
  However, the formal polarisation can be argued to
give a more defined information for a given surface or 
interface, even an absolute value, as was already
proposed by Vanderbilt and King-Smith~\cite{Vanderbilt1993}.


  A compelling piece of evidence in this direction is 
given by the case of the (001) interface between
e.g. KTaO$_3$ (KTO) and LaAlO$_3$.
  Taking the formal charges of -2$e$ for O, +1$e$ for K and
+5$e$ for Ta the (001) layers in KTO 
are +1$e$ (TaO$_2$) and -1$e$ (KO) per layer formula unit. 
  Both LAO and KTO are of the $P_0/2$ kind, and thus one
can foresee a non-polar interface.
  Yet, an ideal epitaxial interface, if considering 
formal charges, would show two $-1$ layers 
or two $+1$ layers together at the interface, since
the $A$O planes are $+1$ for LAO, but $-1$ for
KTO, and vice-versa for the $B$O$_2$ planes.
  Such interfaces would give a net chemical charge
of $\pm 1 \, e/\Theta$, i.e. $\Delta P = \pm P_0$.
  In other words, although an interface between
KTO and LAO could be made without polar 
discontinuity, it seems the ideal epitaxial one
is polar.
  Enhanced carrier densities in
interfaces between III-III and I-V perovskites
have indeed been predicted in ref.~\cite{Cooper2012}.

  The fact is that when defining a whole system
with explicit surfaces, the absolute polarisation 
is well defined, and can be measured, as the
total dipole over the volume of the sample.
  The surface theorem essentially says  
that for specific surface terminations (and as long
as the system remains insulating) the total 
polarisation obtained from the Wannier charge
centres gives the absolute value of the 
polarisation.
  The absolute value of the polarisation is one
specific vector from within the lattice discussed 
in sections~\ref{polarisation} and~\ref{interface planes}. 
  This is true as long as the periodicity parallel to each surface
is respected.
This should be understood in a strict sense: if any local source of 
charge (ion) is added or removed from the surface, it is done 
periodically, and therefore respecting charge quantisation per unit cell.
  This also holds for interfaces. 
  In the KTO/LAO example,
the (001) epitaxial interface, i.e., the one
respecting the $A$O-$B$O$_2$ plane 
alternation, gives a polar discontinuity of
one whole quantum of polarisation.
 However if we were to grow an interface in which 
the first $B$O$_2$ plane of KTO, that is, 
the TaO$_2$ plane, would grow on a
$B$O$_2$ plane of LAO (the AlO$_2$
plane), there would be no polar discontinuity.
  That is, the termination defines the point
in the polarisation set.

\subsection{Alternative formulations}

  Polar surfaces and interfaces have been
known for many decades. 
  The description we have reviewed so far
started in a seminal paper~\cite{Vanderbilt1993}
and is linked to the modern theory of 
polarisation~\cite{Kingsmith1993} but
there were previous characterisations of
such systems.
  The review of Goniakowski, Finocchi and 
Noguera~\cite{Goniakowski2008} gives
an excellent account of the varied descriptions
of polar surfaces and interfaces.
  For classical point particles one finds 
these concepts discussed for related situations
in the book by Born and Huang~\cite{bookBornHuang}.
  Tasker, in 1979, offered a characterisation and
classification of ideally ionic surfaces~\cite{Tasker1979}.
  The link between Tasker's characterisation
and the ideas based on the modern theory of polarisation
was recently established by Stengel~\cite{Stengel2011c}.

  The treatment for quantum delocalised electrons
presents difficulties for some theoretical frameworks, 
however, related to the problem of how to assign a 
static charge to the different atoms in a solid.
  This is the case for instance of the electron 
counting models of, e.g., Harrison and 
coworkers~\cite{Harrison1978} by which
excess charges were assigned to the cations
and anions of heteropolar semiconductors,
trying to account for covalency effects. 
  Such techniques were quite successful at
a qualitative level when used in conjunction
with empirical studies of the electronic structure
of materials. 
  The quantitative determination 
(both from first principles and experimentally) 
of such quantities however is more problematic, since
it involves the assignment of charge to the
atoms in the solid, a badly posed problem.
  It was attacked with different schemes defining
atomic charges, probably the most popular one
in recent times being the one by 
Bader~\cite{BaderCharges}.
  For a brief discussion of the ill-definition
of static charge schemes
in the context of perovskites we refer the reader 
to ref.~\cite{Ghosez1998}.
  
  Interestingly, in highly symmetric systems,
like cubic LAO and STO, such charge redistribution
terms cancel and the final result for the polar
discontinuity coincides with the $P_0/2$ value
established above. 
  For systems in which polarisation is not defined
by symmetry, it is by no means obvious to us 
that the calculation of the polarisation
discontinuity by means of such charge attributions
should agree with the results obtained from 
the calculation of the formal polarisation. 
  We will not discuss them further here, and
we refer the reader to the aforementioned 
review for a quite comprehensive discussion
of such methods~\cite{Goniakowski2008}.

  Finally, another take on the matter worth
mentioning is the one due to Finnis~\cite{Finnis1998}
who discusses polar surfaces and interfaces in
terms of the theory of thermodynamic excesses,
a way of thermodynamically considering 
relative proportions of the different components
in a phase when close to a surface, interface
or grain boundary. 
  Such excesses (changes of concentration
with respect to the bulk phase) also relate
to charge excess, which connects with the
chemical or compositional charge discussed
in section~\ref{net charge}.

\subsection{Electrostatics \label{electrostatics}}

  Within the macroscopic electrostatics relevant 
for length scales larger than atomic size, electric fields 
and charges are related through Gauss's law, 
$\vec \nabla \cdot \vec D = \rho$, where $\vec D$ 
is the electric displacement field, and $\rho$ is the
charge density of free carriers.
  For an interface it becomes, 
\begin{equation}
D^L_z-D^R_z=\sigma,
\end{equation} 
where $\sigma$ is the area density of free charges associated 
to the interface, $z$ is taken as the direction normal 
to the interface and $L$ and $R$ indicate the materials 
on either side of the interface. 
  From the definition of the electric displacement field, 
$\vec{D}=\epsilon_0\vec{\cal E}+\vec{P}$ where 
$\epsilon_0$ is the dielectric permittivity of vacuum, 
$\vec{\cal E}$ the electric field and $\vec{P}$ the 
polarisation, we obtain the change in electric field across 
the interface, 
\begin{equation}
\epsilon_0({\cal E}^L_z-{\cal E}^R_z)=
\sigma-(P^L_z-P^R_z), 
\end{equation}
or
\begin{equation}
\epsilon_0\Delta{\cal E}_z=\sigma-\Delta P_z.
\end{equation}
  Therefore, as stated in section~\ref{net charge}, for electrostatic purposes 
it is only the \textit{sum} of the interface charge and 
the polarisation discontinuity across the interface that 
is required to obtain the resultant electric fields. 
  Separating the spontaneous component from the electric 
field response component of the polarisation, we write 
\begin{equation}
\Delta P_z=\Delta P^0_z + \Delta P_z({\cal E}),
\end{equation}
and, as argued above for LAO-STO,
\begin{equation} 
\sigma - \Delta P^0_z = P_0 / 2 = 0.5 \, e/\Theta,
\end{equation} 
irrespective of whether we use the effective or formal 
polarisation to define the problem, which from now on 
we denote by $\sigma_c$.

  The only thing that remains is to define the geometry 
and electrostatic boundary conditions and to solve for 
the electric fields self-consistently including the dielectric 
responses.
  The most widely studied LAO-STO geometry is a single 
thin film of LAO of thickness $d$ on a TiO$_2$ terminated 
STO substrate~\footnote{For the analogous discussion 
for a superlattice geometry see ref.~\cite{Bristowe2009}.}. 
  This amounts to a system of two interfaces with equal and 
opposite charges, similar to the electrostatics of a capacitor.
  Assuming open-circuit boundary conditions ($\vec{\cal E}
=0$ within the substrate and vacuum), and taking a linear 
dielectric response for LAO, 
\begin{equation}
P_z({\cal E})_{LAO}=\epsilon_0
\chi_{LAO}{\cal E}_{LAO}, 
\end{equation}
we obtain the usual capacitor expression
\begin{equation}
\label{Efield}
{\cal E}_{LAO}=\sigma_c/\epsilon
\end{equation}
where 
\begin{equation}
\epsilon = \epsilon_0 (1+\chi_{LAO}).
\end{equation}
  In other words, in absence of electronic or ionic 
reconstructions, a constant non-decaying electric field 
exists in LAO due to the polar interface charge $\sigma_c$, 
which is partly screened by the dielectric response of LAO.
  This dielectric response of LAO in these films has been 
predicted within first principles calculations
~\cite{Ishibashi2008,Pentcheva2008a,Pentcheva2009}, 
and experimentally inferred through surface x-ray diffraction 
(SXRD) measurements~\cite{Pauli2011} and high-angle annular 
dark-field imaging (HAADF)~\cite{Cantoni2012}
which revealed an off-centring 
of the cation positions along $z$ with respect to the ideal 
perovskite structure,
and through the electrostriction effect~\cite{Cancellieri2011}.


\begin{figure}[t]
\begin{center}
\includegraphics[width=0.7\textwidth]{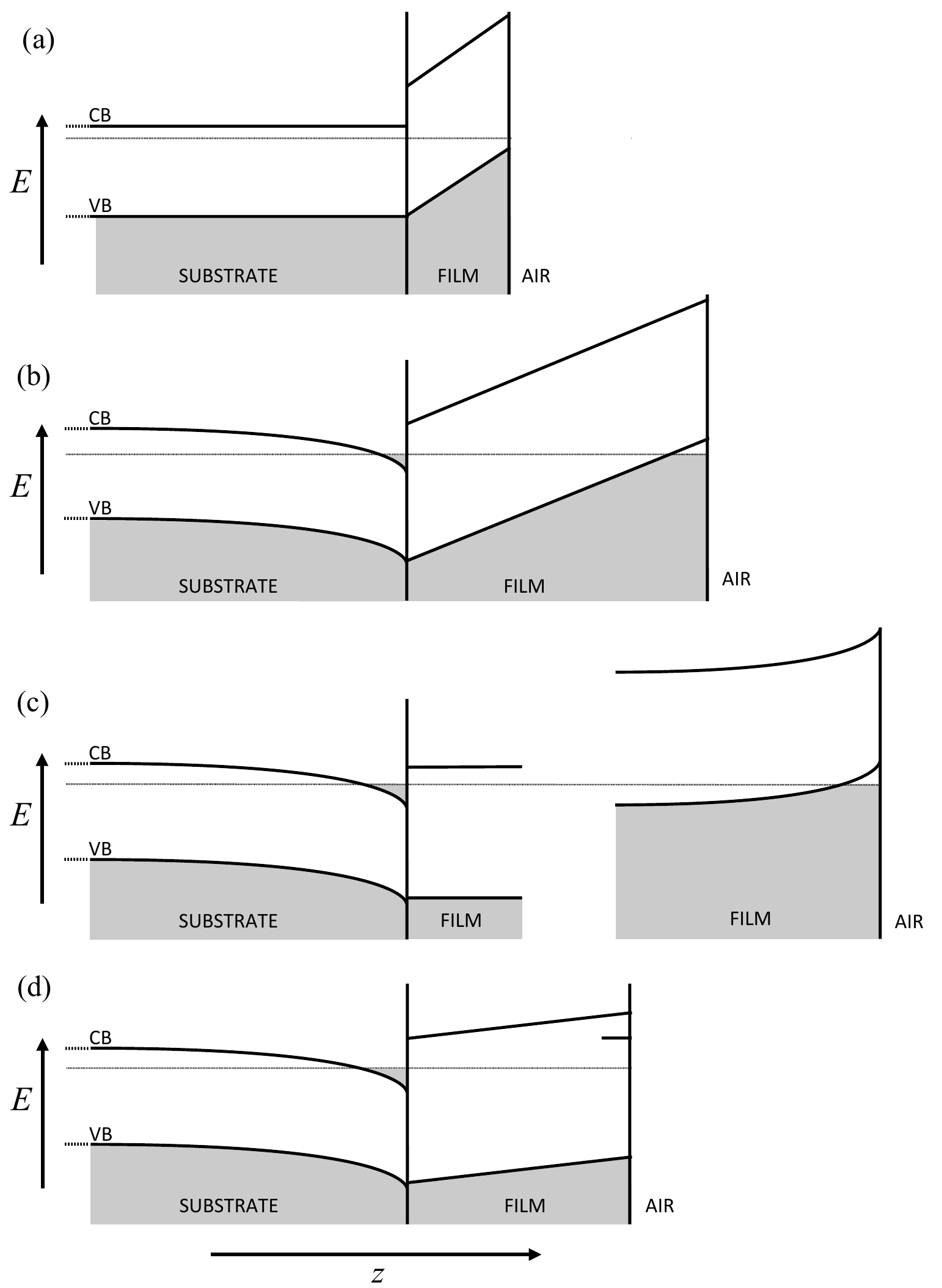}
\caption{\label{BANDS}{
Schematic band diagram of an interface 
between a polar thin film and a thick non-polar substrate, 
adapted from ref.~\cite{Bristowe2011a}. 
  (a) The pristine system under the critical film thickness. 
  (b) Electronic reconstruction.
  (c) The situation for the thick film limit (interface and surface
well separated).
  (d) The introduction of a donor state at the surface and the 
subsequent electron transfer to the interface, reducing 
the electric field in the film.
The dotted horizontal line represents the Fermi energy.}}
\end{center}
\end{figure}

\section{Electronic reconstruction \label{Ereconstruction}}

\subsection{Concept and model}

  We start with a simple model that describes the 
``electronic reconstruction" that has been proposed as 
a mechanism to counter the electrostatic build up that 
originates when growing a thin film of LaAlO$_3$ on 
a SrTiO$_3$ substrate (LAO on STO) along the [001] 
direction~\cite{Nakagawa2006}.
  The model captures the essential physics of the pristine system 
as confirmed by several first principles calculations 
(see for example refs.~\cite{Cen2008,Ishibashi2008,
Lee2008,Pentcheva2009,Bristowe2009}).
  It is illustrated in Figs.~\ref{BANDS} (a) and (b): the 0.5 $e/\Theta$ 
of chemical or compositional charge at the interface and 
surface (corresponding to the half a quantum of formal 
polarisation of LAO, see section~\ref{net charge} and 
Ref.~\cite{Bristowe2011}) gives rise to an electric field 
in the film (see section~\ref{electrostatics}).
This induces a transfer of electrons 
from the valence band at the surface, to the conduction 
band at the interface (see Fig.~\ref{BANDS} b).
The electron transfer only happens once
the potential drop across LAO, $V={\cal E} d$,
reaches the relevant 
band gap, $e\Delta$
(here we will take the gap parameter $\Delta$ as an 
electrostatic potential, so that the relevant band 
gap energy is $e\Delta$).
In this case the relevant band gap
is that of STO if we assume 
no valence band offset for simplicity.
  This process is identical to a Zener breakdown,
although Zener tunnelling itself is not needed. 
  By substituting the expression for the electric field
in Eq.~\ref{Efield} and rearranging, it is simple
to show~\cite{Bristowe2009,Son2009} this transfer happens 
at a critical thickness for this pristine system $d_c^{(p)}$, 
\begin{equation}
\label{critthick}
d_c^{(p)} = \epsilon \Delta / \sigma_c ,
\end{equation}
where $\epsilon$ is the static dielectric constant of the 
film material (LAO).
  Beyond this thickness ($d>d_c^{(p)}$) the electron and hole 
carrier densities reach equilibrium, $\sigma^{(eq)}$, 
pinning the potential drop to $\Delta$,
altering the electric field across LAO, 

\begin{equation}
\label{elao}
{\cal E}_{LAO}=\Delta/d=(\sigma_c-\sigma^{(eq)})
/\epsilon. 
\end{equation}
Rearranging this equation leads to the familiar 
expression for the equilibrium
carrier density,
\begin{equation}
\label{eqdensity}
\sigma^{(eq)} = \sigma_c - {\epsilon
\Delta \over d} = \sigma_c (1-d_c^{(p)}/d).
\end{equation}

  While the critical thickness is not affected, we must 
note here that the last two equations assume 
the strict pinning of the voltage drop to the value of
the gap $\Delta$ independent of carrier density.
  Essentially we are neglecting the kinetic energy of
the electrons in the 2DEG by assuming the limit of a large 
density of states associated to the gas, and thus a negligible
increase in the Fermi level with growing carrier density.
  This has been argued and used in the past in similar
contexts (see e.g. refs.~\cite{Zhu1995,Zhu1996}),
  and is also confirmed by first-principles calculations that
properly include the finite density of states at the interface
(see e.g. ref~\cite{Cancellieri2011}).  

  The consideration of a finite density of states, however, 
does affect the results by reducing the carrier density 
as a function of thickness. 
  Such reduction is zero at $d_c^{(p)}$ but grows with $d$.
It is an important point considering (see section~\ref{Support}) that 
the carrier densities observed experimentally are substantially
lower than expected. 
  It is unlikely, however, that the mentioned effect accounts
for the whole of the discrepancy.
  In fact, at the LAO-STO interface, first-principles calculations
which more properly include these effects produce a sheet carrier
density comparable to that predicted by the initial model
neglecting the kinetic energy of the 2DEG  
(see e.g. ref~\cite{Cancellieri2011}).

  The extension is simple.
Eq.~\ref{elao} becomes
\begin{equation}
\label{elao2}
{\cal E}_{LAO}={\Delta + \sigma^{(eq)}/g\over d}
={\sigma_c-\sigma^{(eq)} \over \epsilon} ,
\end{equation}
for a finite density of states $g$ for the 2DEG (assuming parabolic 
2D bands, the density of states is a constant, independent of energy).
  Solving for $\sigma^{(eq)}$ gives
\begin{equation}
\label{sigma-eq-D}
\sigma^{(eq)} = {\sigma_c d - \epsilon \Delta \over \epsilon/g +d} \, .
\end{equation}
  The critical thickness remains the same, but the curve is
depressed, tending to the same asymptotic value at large
thickness ($\sigma^{(eq)} \rightarrow \sigma_c$), but the
thickness for which $\sigma^{(eq)} = \sigma_c/2$ is larger
by $\delta d = \epsilon/g$.

  Reintroducing the value of $\sigma^{(eq)}$ into Eq.~\ref{elao2},
the thickness dependence of the field is obtained as
\begin{equation}
{\cal E}_{LAO} = {\sigma_c + g \Delta \over \epsilon + g d} \, ,
\end{equation}
connecting with the model and analysis of
ref.~\cite{Bristowe2009}.
  It is no longer pinned to $\Delta/d$.
  
  We introduced $g$ in Eq.~\ref{elao2} as the density of states of
the 2DEG.
  This assumes that the holes at the surface are not mobile.
  If that is not the case, and we do have a 2DEG at the interface 
and a 2D hole gas at the surface, the model is still valid as is,
taking $g$ as the reduced density of states
\begin{equation}
g = {g_e g_h \over g_e + g_h}
\end{equation}
being $g_e$ and $g_h$ the density of states for electrons 
and holes, respectively. That is, $g$ is the density of states
corresponding to the reduced mass of electrons and holes.


\subsection{Model reformulation }

  This model can be reformulated by 
quantifying the formation energy $\Omega_e$ of an area density 
$\sigma$ of electron-hole pairs as they are separated 
by the film's electric field.
The reason for this reformulation will become clear in
section~\ref{sec: redox} when considering the additional
effects of redox processes.
  We account for the competition between two terms;
the energy cost associated with the formation of
an electron-hole pair in absence of the field,
and the energy released 
from the charge compensation of the chemical charge by the
transferred carriers (energy released by 
partly discharging a capacitor), 
as follows:
\begin{equation} 
\label{emodel}
\Omega_e (\sigma) = \left ( \sigma \Delta + \frac{1}{2} \frac{\sigma^2}{g} 
\right ) - \frac{d}{2\epsilon}  \left [ \sigma_c^2 - (\sigma_c - \sigma)^2 \right ] ,
\end{equation}
  The energy is taken with respect to $\sigma = 0$. 
\footnote{A mean-field (Hartree) electron-electron term is 
not explicitly included because the mean-field electrostatic 
interaction is already included in the second term.}

  The equilibrium carrier density $\sigma^{(eq)}$ is then obtained
from minimising $\Omega_e$ with respect to $\sigma$, with the
condition that $\sigma \ge 0$,
\begin{equation}
{\partial \Omega_e \over \partial \sigma} = 
\Delta + {\sigma \over g} + {d \over \epsilon} (\sigma -\sigma_c) = 0 \, ,
\end{equation}
which recovers Eq.~\ref{sigma-eq-D},
as long as $d$ is beyond a critical thickness 
(when $\sigma^{(eq)}=0$)
for this pristine system, 
\begin{equation}
d_c^{(p)} = {\epsilon \Delta \over \sigma_c} ,
\end{equation}
recovering the usual expression 
(Eq.~\ref{critthick}).

  For simplicity, in the remainder of this paper (unless stated) we will
consider the limit of large $g$,
thus also recovering the usual expression for $\sigma^{(eq)}$
(Eq.~\ref{eqdensity})
\begin{equation}
\sigma^{(eq)} = \sigma_c - {\epsilon \Delta \over d} \, .
\end{equation}

  As mentioned in the introduction, the concepts explained
and the model here refer to the final equilibrium situation, not 
to the process of electrons tunnelling from the surface to the
interface. 
  Such process can be quite involved since the situation evolves
while the film is being grown, and it is clearly beyond the scope 
of this review. 
  It is clear, however, that it would be naive visualising the ideal 
process of having grown a film above the critical thickness and
then having the electrons tunnel through at a later stage. 
  It is quite healthy then to consider the point of view of Janotti
{\it et al.}~\cite{Janotti2012}, which, coming from the field
of semiconductor heterostructures, start from the standpoint
that an isolated neutral interface (or surface) of the kind 
we have here, has free carriers compensating for $\sigma_c$.
  Consider an $n$ interface and a $p$ surface coming together
from an infinite distance (film thickness).
  The carriers in both (electrons at the interface, holes at the surface), starting
from the perfectly screened situation, as in Fig.~\ref{BANDS}(c), 
will start to annihilate, equilibrating the Fermi level as in 
Fig.~\ref{BANDS}(b).
 Eventually, for enough proximity (the critical 
thickness) the annihilation is complete. 
  The electric field across the film appears when the carriers
start annihilating, increasing as the thickness is reduced until
it saturates when all carriers have disappeared at (and below)
the critical thickness. 
  Both viewpoints are equivalent in the state they describe for
any thickness, although the rational referring to the process 
is opposite. 
  The model and equations above are equally valid for this
viewpoint. 

  Before we move on from this discussion, we note that the models 
formulated within this review consider a bare LAO thin film on an 
STO substrate under open circuit electrical boundary conditions. 
  The effect of a capping metal electrode on top of LAO has been examined 
from first principles both under open-circuit~\cite{Arras2012,Pentcheva2012}
and finite-$D$~\cite{Cazorla2012} electrical boundary conditions, and 
experimentally under closed circuit and finite-$V$ electrical boundary 
conditions~\cite{SinghBhalla2010}. 
  We simply note here that clearly the picture is altered for these geometries 
and electrical conditions, with the top electrode providing an alternative 
source of electrons. 
Further discussion is beyond the scope of this review, and we refer 
the interested reader to Refs.~\cite{Arras2012,Pentcheva2012,Cazorla2012,SinghBhalla2010}.

\subsection{Support for the electronic reconstruction 
idea and otherwise \label{Support} }

  With Equation~\ref{critthick} one can therefore estimate the 
expected critical thickness for the electronic reconstruction 
in the LAO-STO system simply from the chemical interface 
charge of 0.5 $e/\Theta$, the dielectric constant of the LAO 
film and the relevant band gap of the system - the band gap 
of STO plus the interface valence band offset, and possible
acceptor/donor states at the interface/surface.
  To convert this estimated critical thickness to units of LAO 
unit cells, we also need the out of plane LAO lattice parameter 
when strained to STO. 
  Several values have been quoted in the literature for each 
of these properties.
 Perhaps the most sensitive value being 
the LAO dielectric constant which has been suggested to be 
dependent on strain and electric field.
If calculated {\it ab initio}, the dielectric contstant
is also dependent on pseudopotential, basis set and functional. 
  A recent review paper has summarised this 
sensitivity~\cite{Chen2010}, and the resulting variation in 
estimated critical thicknesses amounts to anything from 4 
unit cells~\cite{Lee2008,Delugas2011,Cancellieri2011} to  6 unit cells or 
even greater~\cite{Chen2009,Pentcheva2009} being 
reported.   


  Perhaps the strongest support for the electronic 
reconstruction model is the fact that experimentally 
a critical thickness of 4 unit cells of LAO has been 
consistently reproduced from transport measurements 
with films grown by both pulsed laser deposition 
(PLD)~\cite{Thiel2006} and molecular beam epitaxy
(MBE)~\cite{Segal2009a}, once STO oxygen vacancies 
are removed by growth in sufficiently high oxygen partial 
pressure and annealing in oxygen~\cite{Basletic2008,
Cancellieri2010}.
 Consistent with these findings is the observation of
a critical thickness of 4 unit cells for 
the presence of a Fermi-edge signal from soft x-ray photoelectron
spectroscopy~\cite{Cancellieri2013}. 
  A quite convincing recent study~\cite{Reinle-Schmitt2012} 
showed that, in agreement with the electronic reconstruction 
model, the critical thickness can be tuned by altering the 
chemical interface charge, $\sigma_c$, through LAO solid 
solution with a non-polar cubic oxide (STO$_{(1-x)}$LAO$_x$). 
  For such system the formal charges argument gives
$\sigma_c = (x/2) \, e/\Theta$, and from Equation~\ref{critthick} 
the electronic reconstruction model would predict a 1/$x$ 
dependence on the critical thickness $d_c^{(p)}$ of the 
diluted film, which is indeed what was obtained in the 
experiment~\cite{Reinle-Schmitt2012}
\footnote{A 1/$x$ 
dependence on the critical thickness $d_c^{(p)}$
can however also be explained within the redox screening
model in the next section, eq.~\ref{redoxcritthick}.}. 

  This argument would seem to contradict
the discussion above that led to the $\sigma_c = (1/2) 
\, e/\Theta$, result as being protected by symmetry. 
  Assuming random alloying of both the $A$ (Sr and La)
and the $B$ cations (Ti and Al) through the film, in an
average (virtual crystal approximation) view, the 
system would remain centrosymmetric and thus
with $P = P_0/2$.
  The paradox is however resolved by noting that
within the average-alloy contemplated scenario, the
relevant quantum of charge renormalises to $x \, e$,
and $\sigma_c$ is still $P_0/2 = (x/2) \, e/\Theta$, 
consistent with both the symmetry arguments above
and the experiment interpretation of 
ref.~\cite{Reinle-Schmitt2012}.

  Additional experimental support for the electronic 
reconstruction model includes the indirect observation of 
a built in electric field in LAO before the critical thickness 
through the electrostriction effect~\cite{Cancellieri2011}, 
and structural~\cite{Pauli2011} and tunnelling 
measurements~\cite{SinghBhalla2010,Huang2012}.
  However it should be noted that spectroscopic 
measurements have not been able to detect a sizeably 
varying field with thickness~\cite{Segal2009a,Slooten2013,
Berner2013b}.
  Finally another strong support for the electronic 
reconstruction model is the fact that several other polar 
interfaces are also conducting, such as interfaces between 
SrTiO$_3$ and LaTiO$_3$~\cite{Ohtomo2002}, 
LaGaO$_3$~\cite{Perna2010}, LaVO$_3$~\cite{Hotta2007}, 
KTaO$_3$~\cite{Oja2012},
KNbO$_3$~\cite{Oja2012}, NaNbO$_3$~\cite{Oja2012}
and GdTiO$_3$~\cite{Moetakef2011}.

  Not all experimental observations are in direct agreement 
with the electronic reconstruction model, however. 
  Measured electron carrier densities are usually found to be 
approximately one order of magnitude lower than that predicted by 
Eq.~\ref{eqdensity} using $\sigma_c=(1/2)\, e/\Theta$~\cite{Thiel2006}. 
  The discovery of a sizeable density of (trapped) Ti 3$d$ like 
states below the critical thickness (at just 2 LAO unit cells)
~\cite{Sing2009,Berner2010,Fujimori2010,Slooten2013} with 
core-level spectroscopic measurements suggests that the 
breakdown may occur almost immediately.
  Additionally no surface hole carriers have been measured 
by transport~\cite{Thiel2006} except when an STO capping 
layer is added~\cite{Pentcheva2010a,Huijben2012}, and no surface hole 
states have been seen near the Fermi level~\cite{Berner2013a,
Plumb2013}.
  These last findings, however, can be explained if the surface 
states are immobile~\cite{Pentcheva2006,Chen2009}.
  Probably more significant, however, is the apparent disappearance 
of conductivity at any LAO thickness for samples grown at very high 
oxygen partial pressures~\cite{Herranz2007,Kalabukhov2011}.
  It is a puzzling result. 
  We can only speculate that this might be connected with the 
dramatically reduced carrier densities~\cite{Schoofs2011,Sato2013,Dildar2013,
Breckenfeld2013,Warusawithana2013a} (and degree of hydroxylation~\cite{Qiao2011a})
observed at certain La:Al off-stoichiometries, which will be further discussed in 
Section~\ref{sec: interface}.
  Other possible mechanisms proposed based on interface effects
will be considered in that section too.

  In the next section we explore the possibility of an alternative 
source of carriers, namely the creation of surface redox defects, 
which can alleviate the polar catastrophe and explain some of the 
issues mentioned within this section with the purely electronic 
reconstruction mechanism. 
  We model the creation of these processes and discuss them 
as the possible origin of the LAO-STO 2DEG in light of available 
experimental data.

\section{Redox screening \label{sec: redox}}

\subsection{General model}

  We now introduce the possibility of there being surface redox 
reactions that leave immobile surface defects with a net charge, 
and free carriers that can move to the interface (see Fig.~\ref{BANDS}(c)).
  A prototypical example~\cite{Bristowe2011a,Li2009,Li2011a,
Cen2008,Zhong2010a} would be that of oxygen vacancies at 
the surface, whereby
\begin{equation}
{\mathrm O}^{2-} \rightarrow {1\over 2} {\mathrm O}_2 + 
{\mathrm V}_{\mathrm O} + 2 e^- .
\end{equation}
  Since the system with the original O$^{2-}$ was neutral and 
the two electron carriers are free to go, the vacancy, V$_{\mathrm O}$,
 is then left with an effective charge $Q=+2e$.
  If the surface is in contact with the atmosphere and water is
present, one can also think of the redox hydroxylation reaction
\begin{equation}
2 {\mathrm O}^{2-} + {\mathrm H_2}{\mathrm O} \rightarrow 
{1\over 2} {\mathrm O}_2 + 2 {\rm OH}^-  + 2 e^- ,
\end{equation}
in which two defects are formed (the two OH$^-$ groups) and 
two carriers are generated (note that the non-redox hydroxylation process,
${\mathrm H_2}{\mathrm O} \rightarrow {\rm OH}^- + {\rm H}^+ $,
does not $a$ $priori$ affect the electrostatics across the film, except through
small alterations in the surface valence band maximum 
and subsequent closing of the effective gap, $\Delta$~\cite{Li2013}).
  Each centre now has a $Q = + e$.
  The mechanism then generalises to the formation of redox 
defects with $Q=Ze$ at the surface and a number $|Z|$ of free 
carriers at the interface. 
  If $Z>0$, the interface carriers will be electrons and the surface 
defects donors.
  $Z=1$ represents single donors (e.g. the hydroxyl groups above), 
$Z=2$ double donors (the O vacancies).
  This corresponds to an oxidation of the surface.
  The opposite is also possible, a surface reduction, with $Z<0$, 
in which case the surface redox defects will be acceptors, and the 
carriers at the interface, holes, which would be the
stable (favourable) situation in the case of a $p$ interface.
   These redox processes are not dissimilar from the pure electronic 
reconstruction we had in section~\ref{Ereconstruction} (electron-hole pair formation)
except that now the surface carriers are immobile (surface defect) 
and have a charge $Q=Ze$, while the interface carriers are mobile 
electrons or holes.
  We consider $Z>0$ in the figures and examples of what follows.
   
  Let us extend the initial model (Eq.~\ref{emodel}) allowing for the 
formation energy of an area density $n$ of such redox 
defects~\cite{Bristowe2011a}.
 These defects compete with electron-hole pairs
, with density $\sigma$, 
from the electron-transfer mechanism:
\begin{eqnarray}
\label{totalenergy} 
\Omega(n,\sigma) = &C n + {1\over 2} \alpha n^2 + \sigma \Delta +
\frac{d}{2\epsilon}  \left [ (\sigma_c - nQ - \sigma)^2 -  
\sigma_c^2 \right ] .
\end{eqnarray}
This now incorporates the formation energy of an isolated redox 
defect in the absence of a field, $C$, and a defect-defect interaction 
term, accounted for in mean-field as ${1\over 2} \alpha n^2$. 
  As in the previous section, the mean field electrostatics is 
captured by the capacitor discharging term (the last term), and
therefore the ${1\over 2} \alpha n^2$ term describes the mean-field
defect-defect interaction beyond electrostatics (e.g. strain, chemical).
  The capacitor discharging term now incorporates the compensation 
due to both charges, the electron-hole pairs ($\sigma$) and the ones coming 
from the redox processes ($nQ$).
  The free charge density at the interface is now the sum of the two, 
$\sigma_{2DEG}=\sigma + nQ$.
  The $C$ constant depends on the chemistry (the breaking and 
making of chemical bonds) and on the chemical potential of the 
species resulting from the redox process. 
  In the two reactions specified above, the chemical potential of
O$_2$, which depends on its partial pressure in the gas in equilibrium 
with the system, and the temperature.
  The humidity (chemical potential of H$_2$O) will affect the second 
reaction.

\begin{figure}
\begin{center}
\includegraphics[width=0.70\textwidth]{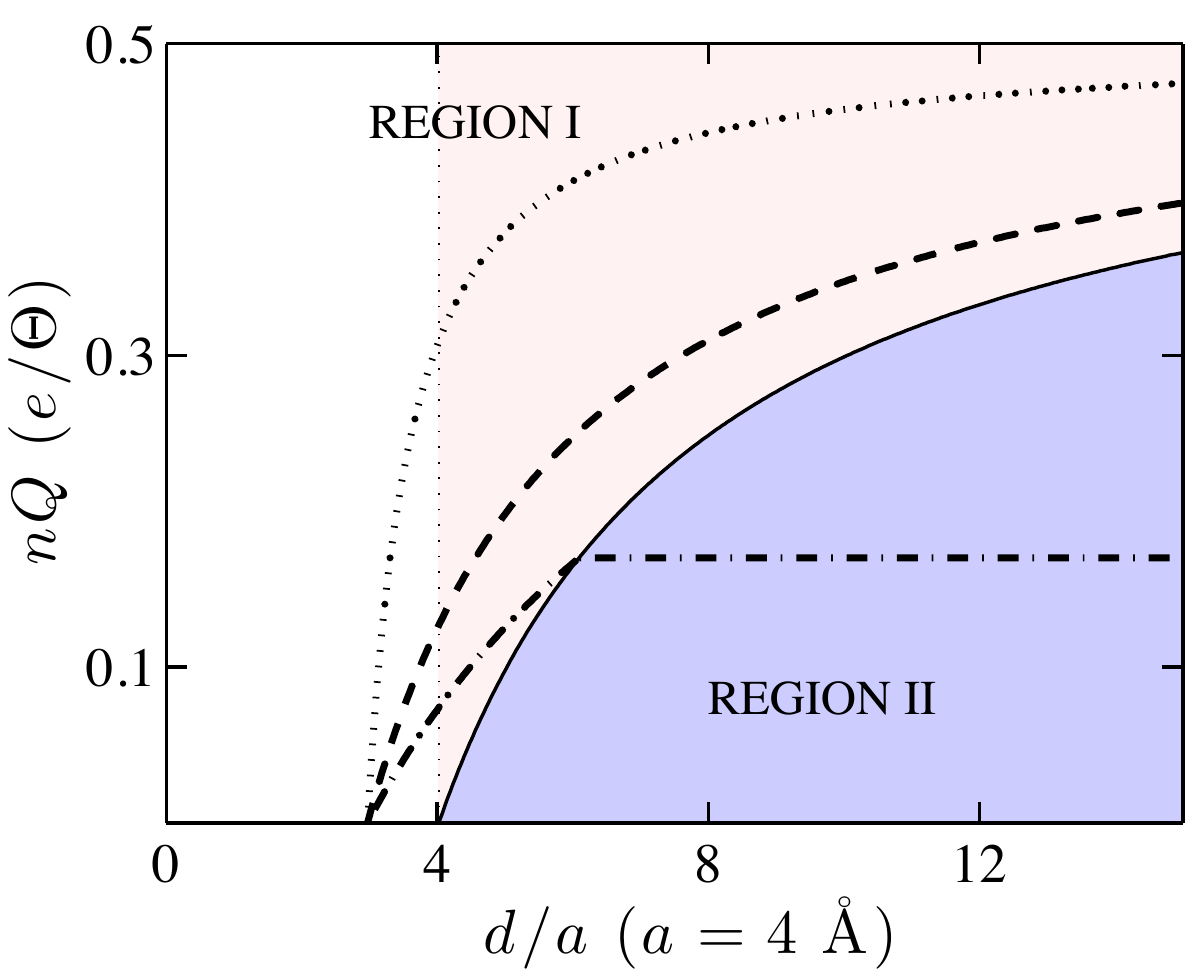}
\caption{\label{PHASES} Equilibrium redox charge $Qn^{(eq)}$ vs
film thickness $d$ for $\sigma_c = 0.5$ $e/\Theta$, $C=4.8$ eV~\cite{Bristowe2011a},
 $\epsilon=28$~\cite{Delugas2005}, $\Delta=3.25$ eV~\cite{VanBenthem2001} and $\alpha=0.8$ 
eV\AA$^4$ (dashed line), -15 eV\AA$^4$ (dotted line) and 
20 eV\AA$^4$ (dash-dotted line).
  The light (dark) region corresponds to $\sigma=0$ ($\sigma > 0$).
  The vertical thin dotted line indicates the discontinuity of
$\tilde \Omega$ described in section~\ref{subsec: fp}.}
\end{center}
\end{figure}

  Again, we find the equilibrium concentrations for $\sigma$ and $n$ 
by minimising $\Omega$, subject to the conditions $n \ge 0$ and 
$\sigma \ge 0$.
\begin{equation}
{\partial \Omega \over \partial \sigma} = 
\Delta + {d \over \epsilon} (\sigma + n Q -\sigma_c) = 0
\end{equation}
implies that
\begin{equation}
\label{sigeq}
\sigma^{(eq)} = \sigma_c - nQ  - {\epsilon \Delta \over d} .
\end{equation}
  The condition $\sigma \ge 0$ implies
\begin{equation}
\label{phaseboundary}
n Q \le  \sigma_c  - {\epsilon \Delta \over d} ,
\end{equation}
meaning that in the plot of $nQ$ vs $d$ (see Fig.~\ref{PHASES}) there is a 
separation of two regions, one with conventional electron-hole formation, 
the other without it.
  As expected, the boundary line hits the $n=0$ line at $d = d_c^{(p)}$, the
critical thickness for the appearance of electron-hole pairs in the absence of
redox defects.
  It is important to note here that the energy $\Omega$ will be continuous 
across the boundary between regions, albeit its derivative will not. 

  Let us now find the equilibrium concentration of surface redox defects.

\subsubsection{ Region I: $\sigma^{(eq)}=0$. } 
  Consider first the region in which there is no 
electronic reconstruction, $\sigma^{(eq)}=0$.
  The energy becomes: 
\begin{equation} 
\Omega(n) = C n + {1\over 2} \alpha n^2 + 
\frac{d}{2\epsilon}  \left [ (\sigma_c - nQ )^2 -  \sigma_c^2 \right ] ,
\end{equation}
and $n^{(eq)}$ is found as ever:
\begin{equation}
\label{neq}
{\partial \Omega \over \partial n} = 0 \;  \;  \;  \; \Rightarrow \; \; \;  \; 
n^{(eq)} = { Q d \sigma_c - \epsilon C \over Q^2 d + \epsilon \alpha}
\end{equation}
%
as already presented in Ref.~\cite{Bristowe2011a}.
  In this case, there is a critical thickness for the onset of redox screening, 
i.e., when $n^{(eq)} \ge 0$, which happens at the critical thickness
\begin{equation}\label{dcLAO}
\label{redoxcritthick}
d_c = {\epsilon C \over Q \sigma_c} .
\end{equation}
which is relevant only while in the $\sigma=0$ regime.
  That means that  $d_c$ is meaningful as such if smaller than $d_c^{(p)}$.
  Using their definitions, this happens when $C \le Q \Delta$.
  In practice this corresponds to cases where the redox process is
energetically more favoured than the creation of electron-hole 
pairs in zero-field.

  The $\alpha$ parameter is expected to be small~\cite{Bristowe2011a},
therefore let us see what happens for $\alpha=0$.
  Then
\begin{equation}
n^{(eq)} Q =  \sigma_c - {\epsilon C \over Q d} \: ,
\end{equation}
which, if compared with Eq.~\ref{phaseboundary}, shows
that the equilibrium redox defect density runs parallel to the 
boundary between the two regions ($\sigma^{(eq)}=0$ and 
$\sigma^{(eq)} > 0$), asymptotically as $1/d$ towards $\sigma_c$.

  For $\alpha > 0$, the $n^{(eq)}$ onset is still the same, at $d_c$, and
the curve still tends asymptotically to $\sigma_c$ as $1/d$.
  However, the behaviour in between is not $1/d$, and the value 
of $\alpha$ pushes the curve down and makes it cross the boundary 
at some thickness $d_x$. 
%
%
%
%
  Such crossing can be obtained by finding the thickness for which 
$n^{(eq)}$ equals the boundary density, Eq.~\ref{phaseboundary}, 
\begin{equation}
\label{neqcrossing}
n^{(eq)} = {\sigma_c \over Q} - {\epsilon \Delta \over Q d_x} \; ,
\end{equation}
which, using Eq.~\ref{neq}, gives
\begin{equation}
\label{dx}
d_x = { \epsilon \Delta \alpha \over Q C - Q^2 \Delta + \alpha \sigma_c} \; .
\end{equation}
This last equation becomes clearer when written as follows
\begin{equation}
d_x = d_c^{(p)} { d_{\alpha} \over d_{\alpha} - \delta d } \; ,
\end{equation}
where 
\begin{equation}
d_{\alpha} \equiv {\epsilon \over Q^2} \alpha \; \; \; {\rm and} \; \; \; 
\delta d \equiv d_c^{(p)} - d_c \; .
\end{equation}
  For the crossing we seek, $d_x > d_c^{(p)}$, we then need that
$d_{\alpha} >  d_{\alpha} - \delta d$ (which happens whenever
$d_c < d_c^{(p)}$), and we need the denominator to be positive, i.e., 
\begin{equation}
\label{crossing}
d_{\alpha} > \delta d \; .
\end{equation}
  This means that if the onset for redox defect stabilisation happens 
before the onset for electron-hole pair formation, the equilibrium 
line will remain in region I (no additional electron-hole formation, 
$\sigma=0$) for all thicknesses, unless the surface defect mutual 
repulsion is strong enough.
  In this case of $\sigma=0$ the free charge at the interface is 
simply $n^{(eq)}Q$ and can reach $\sigma_c$ rather quickly.
  It is interesting to briefly note here a comparison with the
experimental results in ref.~\cite{Cancellieri2011}, where
complete screening of the electric field was inferred through 
the electrostriction effect for 
films as thin as 6 LAO perovskite layers. 
   
  It remains to be seen what happens if the equilibrium 
line crosses to region II, 
or if the onset was already in region II, or, even if there is an 
equilibrium line always in region I, whether there may be another 
in region II.
  Before we get there, let us find the value of $n^{(eq)}$ at the 
crossing point.
  For that we introduce the value of $d_x$ of Eq.~\ref{dx} into 
$n^{(eq)}$ as in Eq.~\ref{neqcrossing}, 
\begin{equation}
n^{(eq)} (d_x) = {\sigma_c \over Q} - {\epsilon \Delta \over Q}
{q C - Q^2 \Delta + \alpha \sigma_c \over \epsilon \alpha \Delta}, 
\end{equation}
which yields
\begin{equation}
\label{neqbound}
n^{(eq)} (d_x) = {Q \Delta - C \over \alpha} \; ,
\end{equation}
which is below the asymptotic limit of $\sigma_c / Q$ if $\delta d 
< d_{\alpha}$, and hence consistent with the previous discussion 
leading to Eq~\ref{crossing}.

\subsubsection{Region II: $\sigma^{(eq)} > 0$}

  In this region the equilibrium electron-hole planar density 
$\sigma^{(eq)}$ is the one given by Eq.~\ref{sigeq} in the previous 
section.
  We now find the equilibrium density of redox defects by 
minimising $\Omega$ (in Eq.~\ref{totalenergy}) with respect to $n$:
\begin{equation}
{\partial \Omega \over \partial n} = C  + \alpha n - {d \over \epsilon}  
(\sigma_c - n Q - \sigma ) Q  = 0 \; .
\end{equation}  
Introducing $\sigma^{(eq)}$ and solving, we obtain
\begin{equation}
\label{neq2}
n^{(eq)} = {Q \Delta - C \over \alpha} \; ,
\end{equation}
which is thickness independent and coincides with the boundary 
value of $n^{(eq)}(d_x)$ coming from region I (see 
Eq.~\ref{neqbound}).
  This means that if $n^{(eq)}$ of region I crosses the boundary, 
it continues horizontally into region II.
  If, on the other hand it does not cross, then there is no minimum 
for $n$ in region II (if $d_{\alpha} < \delta d$, then $( Q \delta - C) 
/ \alpha$ gives a value beyond $\sigma_c / Q$), and the complete 
solution found in region I remains unique.

  Finally, if the onset of defect formation happens for thicker 
samples than the onset for electron transfer, i.e. $d_c > d_c^{(p)}$, 
then the equilibrium $n$ in region II becomes negative for all 
thicknesses, meaning that $n^{(eq)} = 0$ always, i.e. no redox 
defects appear.
In this case we recover the solution of the previous section, 
$\sigma_{2DEG}=\sigma$.

\subsection{Comparison with first-principles calculations 
\label{subsec: fp}}

\subsubsection{Formation energy of redox defects}

  The energy $\Omega$ within sections~\ref{Ereconstruction} and~\ref{sec: redox} 
accounts for the formation 
energy of both redox defects and electron-hole pairs.
  Sometimes there is interest in the energies themselves (for instance, 
if comparing with electronic structure calculations).
 Depending on what is being compared, the relevant energy may be the 
energy of the system with redox defects referred to the energy 
the system would have if only allowing for electronic transfer. 
  In that case the relevant energy is
\begin{equation}
\tilde \Omega(n) =   \Omega(n,\sigma^{(eq)}) - 
                                   \Omega_e(\sigma_e^{(eq)}) \; .
\end{equation}

  It does not affect the equilibrium concentrations for $n$,
since we are just adding an $n$-independent energy to what 
we had.
  The only difference in the analysis is that there would be a 
discontinuity in the derivative of the energy $\tilde \Omega$ at 
$d_c^{(p)}$ because of there being a kink in $\Omega_e$. 
  It is indicated in Fig.~\ref{PHASES} by a vertical dotted line, 
effectively defining a third region.
  The resulting energy in the different regions is
%
\begin{equation}
\hspace*{-2cm}
\tilde \Omega(n)  =\cases{
 C n + {1\over 2} \alpha n^2 + {d\over 2 \epsilon}
\left [ (\sigma_c - nQ )^2 -  \sigma_c^2 \right ]  &for $d < d_c^{(p)}$\\
C n + {1\over 2} \alpha n^2 + \Delta({\epsilon\Delta\over 2d} -  
\sigma_c)+{d\over 2 \epsilon}(\sigma_c -nQ)^2 &for 
$n Q \ge  \sigma_c  - {\epsilon \Delta \over d} $\\
(C-Q\Delta)n + {1\over 2} \alpha n^2 &for 
$n Q \le  \sigma_c  - {\epsilon \Delta \over d}.$\\}
\end{equation}
  
\subsubsection{Formation energy of one redox defect}

  Given the fact that $\tilde \Omega(n)$ is the energy for a
given density of defects $n$, the formation energy of one 
redox defect in an infinite system with a defect density $n$ 
would be
\begin{equation}
E_f (n) = {\partial \, \tilde \Omega(n) \over \partial n} \; .
\end{equation}
\begin{figure}
\begin{center}
\includegraphics[width=0.50\textwidth]{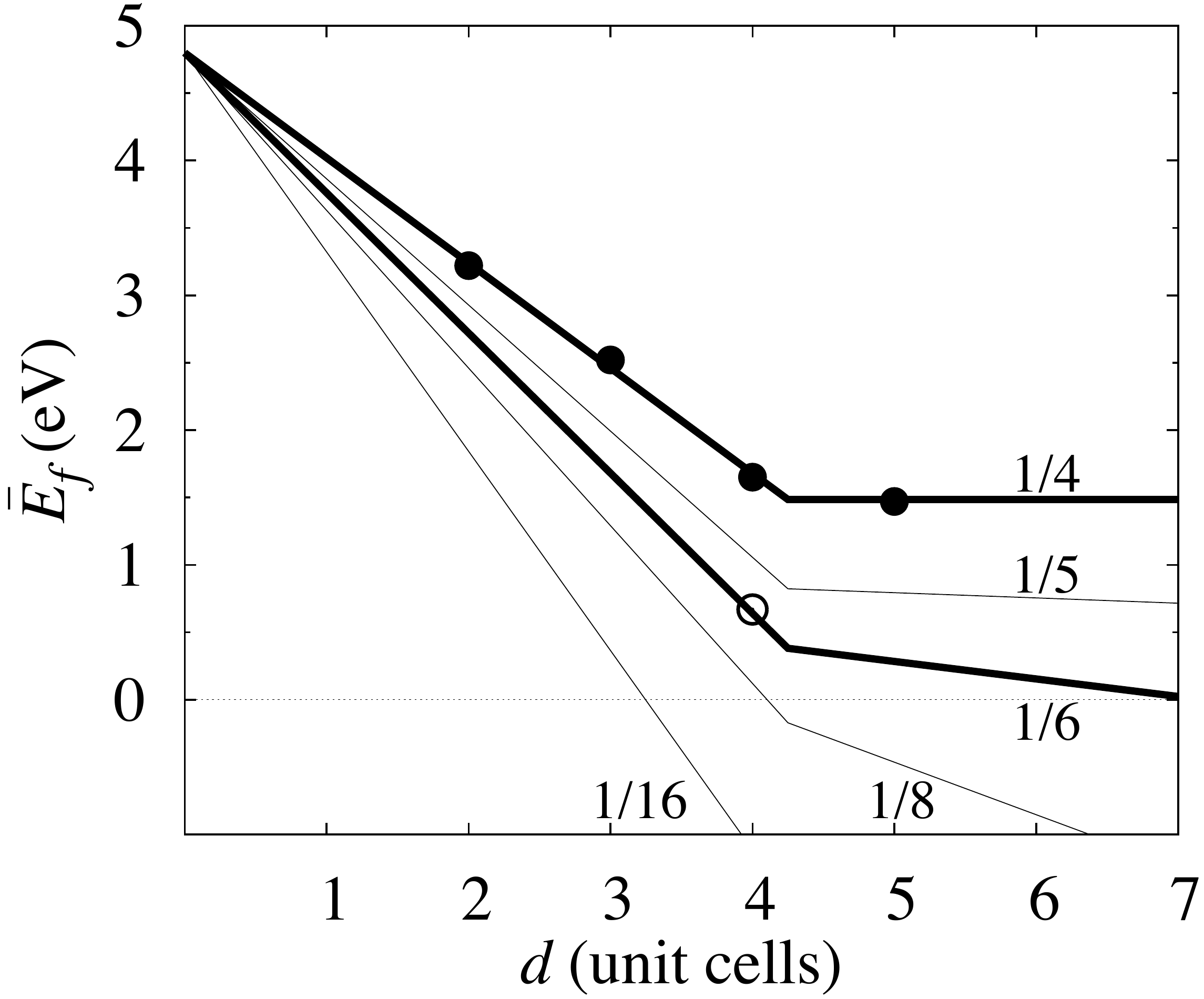}
\caption{\label{E_F} Redox defect formation energy (per defect) 
$\bar E_f$ as a function of thickness $d$ for various values 
of the final redox defect concentration $n$ (adapted from 
ref.~\cite{Bristowe2011a}). 
  The model (lines) is compared with first principles 
(points)~\cite{Li2009,Li2011a}.}
\end{center}
\end{figure}
  However, to compute the formation energy of one redox defect 
from first-principles one would calculate the difference in energy 
between a thin film with one redox defect in a given supercell, 
which defines the concentration $n$, and the energy of the 
same supercell without the defect.
  That quantity we will call $\bar E_f$, and relates to the previous 
magnitudes in the following way:
\begin{equation}
\bar E_f (n) = {\tilde \Omega (n) \over n}  = {1\over n} \int_0^n \! 
E_f(n') {\mathrm d}n' \; , 
\end{equation}
which results in 
\begin{equation}
\hspace*{-2cm}
\bar E_f(n)  =\cases{
 C + {1\over 2} \alpha n + {d\over 2 \epsilon n}
\left [ (\sigma_c - nQ )^2 -  \sigma_c^2 \right ]  &for $d < d_c^{(p)}$\\
C + {1\over 2} \alpha n + {\Delta \over n}({\epsilon\Delta\over 2d} - 
 \sigma_c)+ {d\over 2 \epsilon n}(\sigma_c -nQ)^2 &for 
 $n Q \ge  \sigma_c  - {\epsilon \Delta \over d} $\\
C-Q\Delta + {1\over 2} \alpha n &for 
$n Q \le  \sigma_c  - {\epsilon \Delta \over d} .$\\} 
\end{equation}
 
  Figure~\ref{E_F} shows the $\bar E_f$ for various concentrations,
as a function of the thickness $d$, where it is compared with several 
surface vacancy formation energies obtained from first 
principles~\cite{Li2009,Li2011a}.
  The agreement is remarkable taking into account that the 
parameters could be independently computed ($C$ is hard to pin
in comparisons with experiments since it contains the chemical
potential, which is hard to establish; but it is easier in comparison
with calculations where the energy of the reference structures is
well defined; see the Appendix in Ref.~\cite{Bristowe2011a}).
The formation of surface oxygen vacancies in LAO-STO has been 
further ratified within the careful calculations of ref.~\cite{Yu2014} only for the 
ideal concentration of oxygen vacancies, without considering oxygen 
vacancy concentration dependence.

\subsection{Generalisation}

  The situation described in this section can be seen as that of competing 
screening mechanisms to the depolarising field, both $n$ and $\sigma$
being positive definite, and both with their respective energy thresholds.
  One can think of the general screening of a thin film by different such 
mechanisms by rewriting Eq.~\ref{totalenergy} as
\begin{eqnarray}
\Omega(\{\sigma_i\}) =  &\sum_i \left ( \sigma_i \Delta_i + 
{1\over 2} \alpha_i \sigma_i^2 \right ) 
+ \frac{d}{2\epsilon}  \left [ \left (\sigma_c  - \sum_i \sigma_i \right )^2 
-  \sigma_c^2 \right ] .
\end{eqnarray}
  In the case of electronic reconstruction the quadratic term can
also be included, as $\alpha_i=1/g$, the inverse density of states,
accounting for the kinetic energy of the carriers, as discussed in 
Sections 3.1 and 3.2.


  Similarly, this generalised energy could include bilinear 
couplings in the different variables, as $\sum_{ij} \beta_{ij} 
\sigma_i \sigma_j$. 
  In the case of $\sigma$ and $nQ$, if the energy in 
Eq.~\ref{totalenergy} had an added term $\beta n Q \sigma$, 
the equilibrium electron density (Eqs.~\ref{sigeq}) would read
\begin{equation}
\sigma^{(eq)} = \sigma_c - nQ \left (1 + {\beta \epsilon \over d} \right )  
- {\epsilon \Delta \over d} .
\end{equation}
and the phase boundary (in Eq.~\ref{phaseboundary}) given by the 
condition $\sigma > 0$ would transform into
\begin{equation}
n Q <  {\sigma_c  - {\epsilon \Delta \over d} \over 1 + {\epsilon \beta \over d} } .
\end{equation}
  Such a term would describe the change in effective gap when in the
presence of surface carriers due to the electronic reconstruction.
  It is important to state now, however, that under such interactions
it is possible that the relevant defects would cluster, an effect beyond
the mean-field description implied in the equations. 
  This discussion will not be pursued any further here.

\subsection{Experimental signatures and discussion}

  We conclude this section with a summary of experimental 
evidence for and against the redox screening mechanism.
  We concentrate on the LAO-STO system for which there are 
numerous studies for comparison.  

  The first expectation from redox screening is obviously that 
the surface remains insulating (except for ionic diffusion) 
since it is charge compensated by bound charge centres 
(e.g. oxygen vacancies), whilst the interface is screened by 
electrons (which may or may not be mobile). 
  As already mentioned, hole conduction at the LAO surface 
has not been found for any thickness of LAO, whilst electronic 
conduction appears at the LAO-STO interface at 4 unit cells 
of LAO or more~\cite{Thiel2006}. 
  Ti 3$d$ like states (Ti 3+) have been observed from 
spectroscopic measurements even below the 4 unit cells 
critical thickness, and as thin as just 1 unit cell~\cite{Sing2009,
Berner2010,Fujimori2010,Slooten2013}. 
  This suggests the possibility of the redox screening 
mechanism appearing almost immediately with LAO growth, 
since with the purely electronic reconstruction one would 
expect surface hole conduction and only an electronic 
transfer to the interface once LAO is approximately 4-6 
unit cells thick (see section~\ref{Ereconstruction}).
  It is indeed shown in ref.~\cite{Bristowe2011a} following 
the same recipe as in section~\ref{sec: redox} that the formation of surface oxygen 
vacancies in LAO-STO is likely to occur for LAO films as 
thin as 1-3 unit cells thick. 

  The rapidly increasing interface electron density with 
film thickness predicted within the redox model 
(for a certain range of $\alpha$)
is consistent with
the experimental results in 
ref.~\cite{Cancellieri2011}, where nearly
complete screening of the electric field was inferred through 
the electrostriction effect for 
films as thin as 6 LAO perovskite layers. 
This is too thin to be explained by the 
electronic reconstruction model alone. 

  If redox formation is indeed responsible for the observed 
Ti 3$d$ states, the next question is why for films thinner 
than 4 unit cells are these electrons not available for conduction. 
  We briefly mention three possibilities that come to mind. 
  Firstly it has been suggested that the 2DEG lies in several 
Ti 3$d$ sub-bands, some of which are not mobile due to 
Anderson localisation~\cite{Popovic2008}.
  The other two mechanisms are related to the experimental 
observation that the immobile Ti 3$d$ levels are `in-gap' 
states~\cite{Drera2011,Ristic2012,Koitzsch2011,Berner2013a} 
at higher binding energies. 
  The second possibility is that acceptor defect levels are formed 
during the LAO growth process, such as from cation intermixing 
(see section~\ref{sec: interface}). 
  Cation intermixing has been readily observed in most 
LAO-STO samples~\cite{Nakagawa2006,Chambers2010,
Qiao2010,Qiao2011,Pauli2011,Willmott2007,Gunkel2010,
Kalabukhov2009,Vonk2012}.

  The third possibility is that such states could be formed from 
the point charge of the surface redox defects which generate 
trapping potentials as discussed in ref.~\cite{Bristowe2011a}.
  As the LAO film thickness increases the traps change from 
deep and few to shallow and overlapping.
 Hence a transition from insulating to conducting occurs at larger 
LAO film thickness than the redox formation itself. 
  The character of the transition would depend on the
charge of the redox defects $Z$. 
  For even $Z$ (as in the double-donor case of the oxygen
vacancies), the transtion would be that of band overlap, 
whenever the band of filled states would touch the bottom
of the conduction band.
  For odd $Z$ (as in the case of OH$^-$ defects) the 
partly filled gap states would undergo an insulator to metal
transition of the Mott-Anderson type~\cite{Bristowe2011a}.

  Several other recent experimental findings, some of 
which incompatible with the electronic reconstruction, 
find natural explanation within the redox screening model.
  XPS measurements have found very little, if any, evidence 
for core level broadening or shift with increasing film 
thickness~\cite{Segal2009a,Slooten2013,Berner2013b} at odds with 
what one might expect from a constant electric field within 
LAO before the electronic reconstruction critical thickness.
  Within the redox screening model, it can be trivially 
shown that the potential drop across the LAO film is 
essentially independent of thickness (if $\alpha$ is 
small)~\cite{Bristowe2011a}.
  Using the parameters for LAO-STO the potential drop 
for each LAO layer added was predicted to be between 
0.0 and 0.2 eV~\cite{Bristowe2011a}, much 
smaller than within the electronic reconstruction model 
(approximately 0.7-0.9 eV).  
  The predicted pinning of the potential drop (and hence 
reduction in electric field with thickness) is also consistent 
with the reduced cation-anion buckling with increasing 
LAO thickness as observed by SXRD~\cite{Pauli2011}. 
  
    Finally we note the observation that the field effect switching 
of the LAO-STO conductivity~\cite{Thiel2006,Cen2008} is intimately 
linked with surface charge writing~\cite{Xie,Xie2011a} and 
that the process is dependent on the existence of water in 
the surface environment~\cite{Bi2010}.
 These experiments are likely linked 
with electrochemical processes~\cite{Bark2012,Kumar2012}, 
and are all compatible with the redox screening model.
  Applying a biased tip to the surface alters the field across 
the LAO film which either increases or decreases the stability 
of surface redox defects (and hence interface charge carriers 
and surface charge) depending on the sign of the bias. 
  Such electrochemical processes have also been predicted 
during the switching of ferroelectric thin 
films~\cite{Wang2009,Fong2006,Spanier2006,Stephenson2011,Bristowe2012}.


\section{Interface effects \label{sec: interface}}

  So far we have assumed perfect interfaces {\it and}
thin two-dimensional charge distributions.
  In this section we explore the effect on the phenomenology
of deviations from such assumptions.

\subsection{Interface dipole due to the electrons}

\begin{figure}
\begin{center}
\includegraphics[width=0.30\textwidth]{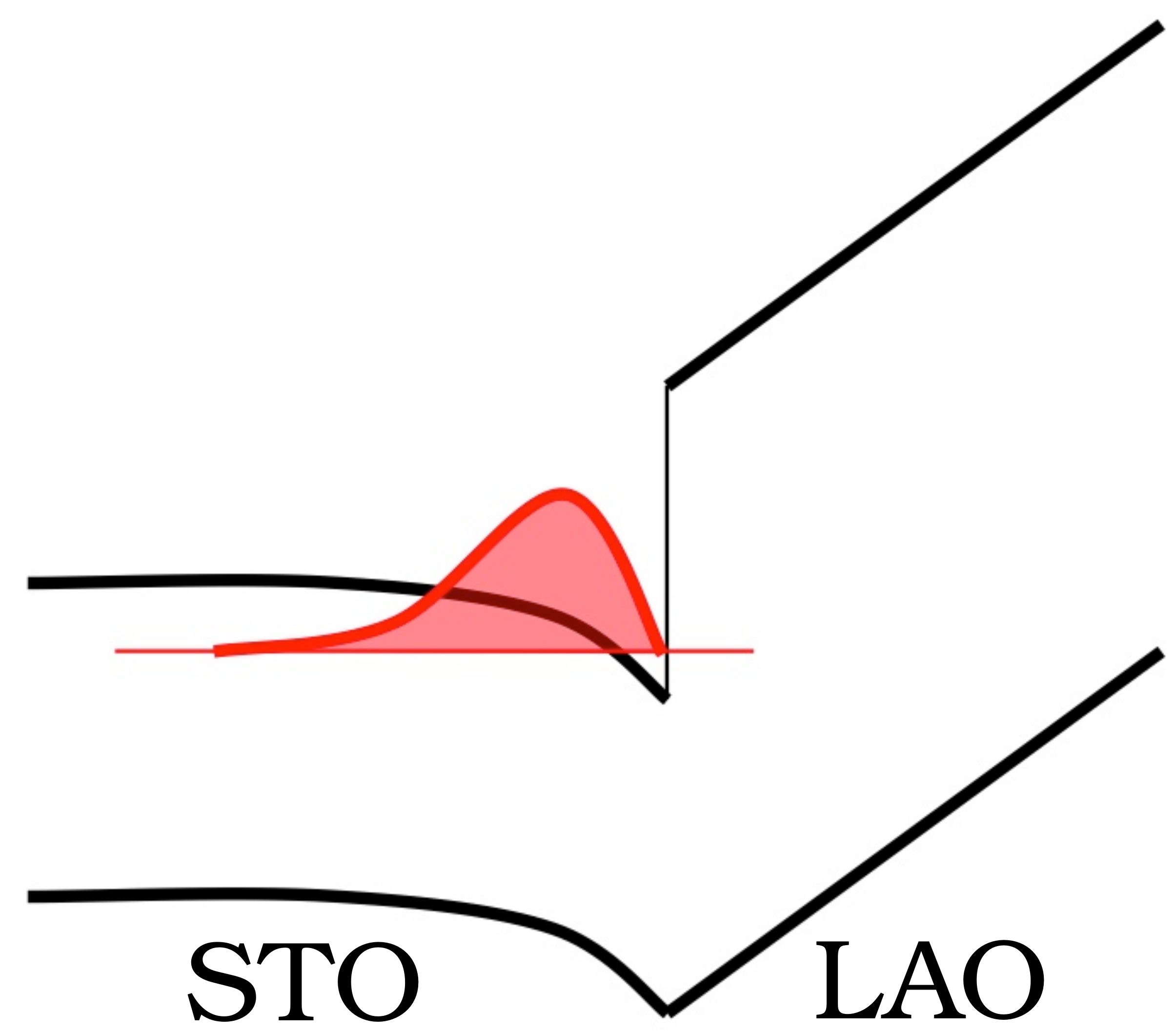}
\caption{\label{sketch} 
Sketch of band gap and interface electron density 
(red) vs distance perpendicular to the interface $z$.}
\end{center}
\end{figure}

  The free carriers at the interface do screen (some of)
the chemical charge at the interface, but their quantum spread 
in $z$ generates an interface dipole that has been proposed
to alter the relative alignments between LAO surface and STO's
bottom of the conduction band. 
  This effect is well described in the literature (see e.g.
ref.~\cite{Stengel2011} and references therein).
  The relevant alignment~\cite{Stengel2011} is 
the one given by the filling of the carrier bands in the 2DEG,
as illustrated in Figure~\ref{sketch}.
  It depends on the zero-point energy 
of the band, and therefore on the perpendicular effective 
mass of the carriers, and on the 2DEG band filling,
which depends on the carrier density and the
parallel effective mass of the carriers.
  Discussions on the character of the 2DEG bands and
corresponding effective masses are beyond the scope
of this review.


\subsection{Interface cation intermixing}

  Both during and after sample growth there is the
possibility of inter-diffusion of atoms across the interface.
  Such processes have attracted attention
lately~\cite{Nakagawa2006,Pauli2011,Chambers2010,Qiao2010,
Qiao2011,Willmott2007,Gunkel2010,Kalabukhov2009} 
as one of the possible explanations for the origin of 2D 
conduction in LAO/STO interfaces~\cite{Schlom2011}.
  The argument used is simply that, since La is a known 
dopant (donor) in STO, La cations diffusing from LAO 
into STO could dope the interface and thus make it 
conducting.

  Let us consider here what to expect from ions 
exchanging across the interface\footnote{In what 
follows we will make two approximations to simplify 
discussions, both of which can be trivially added to 
the models. 
  Firstly we take the pristine interface dipole to be 
zero, and secondly we make the approximation 
that the dielectric constant is unaltered by  inter-diffusion 
across the interface.}.
  This discussion will focus on the effects of having
a given inter-diffusion profile in a sample, rather than the 
energetics and kinetics that might originate it, since
equilibrium is not expected to be reached in this aspect.
  We shall concentrate here on inter-diffusion situations
starting from the pristine stoichiometric interface,
and all the exchanges happening within a region
around the interface whose width is smaller than
other length scales in the problem.
  The effects on deviations from stoichiometry, and of
diffusion processes from/to reservoirs extrinsic to
the interface, are qualitatively different and will be 
treated briefly in section~\ref{stoichiometry}.

\subsubsection{Interface doping generated by inter-diffusion}

  Based on chemical (steric) arguments and on quite a 
wealth of accumulated knowledge on perovskites,
the most likely inter-diffusion processes by far are swaps 
of like atoms, namely, $B$ cations (Al$^{3+}$ cations 
in LAO substituted by Ti$^{4+}$, and Ti$^{4+}$ in STO
substituted by Al$^{3+}$) and $A$ cations (La$^{3+}$ to 
Sr$^{2+}$ on one side, Sr$^{2+}$ to La$^{3+}$ on the
other).
  The swap of O$^{2-}$ anions across the interface is also likely,
but of no consequence.

  Each cation substitution gives rise to a net bound 
charge plus a compensating carrier.
  La$^{3+}$ replacing Sr$^{2+}$ gives a
$+e$ centre with an associated electron, thus
the donor mentioned in the previous subsection, and so do the other
substitutions with obvious sign for their charges.

  The first important point to make here is that such inter-diffusion 
{\it should not be expected to dope the interface}.
  Be it $A$ or $B$ cations swapping across the interface, there will
be as many ``wrong" cations on one side as on the other, and
thus as many donors at one side as acceptors at the other.
  The electrons of the donors annihilate with the holes of
acceptors, as long as both dopants are located in the proximity
of the interface, the situation considered here.
  Similarly, there will be as many Ti$^{4+}$ donors in LAO as
Al$^{3+}$ acceptors in STO.
  The inter-diffusion will then originate a distribution of
fairly immobile charges at both sides of the interface, 
with zero total charge, since there are as many $+e$'s
as $-e$'s. 
  Indeed it is well known that the LAO-STO solid solution is insulating. 

  While intermixing is not expected to dope the interface, 
it can have implications for the \textit{transport} properties
and critical thickness of an interface 2DEG appearing by
means of another mechanism.
  This is likely since the intermixing can produce the 
aforementioned trapping acceptor 
states lying below the conduction band 
of the pristine interface.

\subsubsection{Electric field, interface dipole and interface shift}

  The second important point to make here is that such 
inter-diffusion does affect the electrostatics, but only in 
the interface region.
 It does not affect the electric field 
emanating from it into the film.
   This is a direct consequence of Gauss's law: the net flux 
of the electric displacement field across surfaces parallel 
to the interface equals the net charge contained in the 
region within them.
   In the absence of charges diffusing in from other regions,
the net charge per unit area is still $\sigma_c$, and the
field emanating from the interface into 
the film is again ${\cal E} = \sigma_c  /  \epsilon_{\rm LAO}$, 
(the field in the substrate being zero).
        
\begin{figure}
\begin{center}
\includegraphics[width=0.70\textwidth]{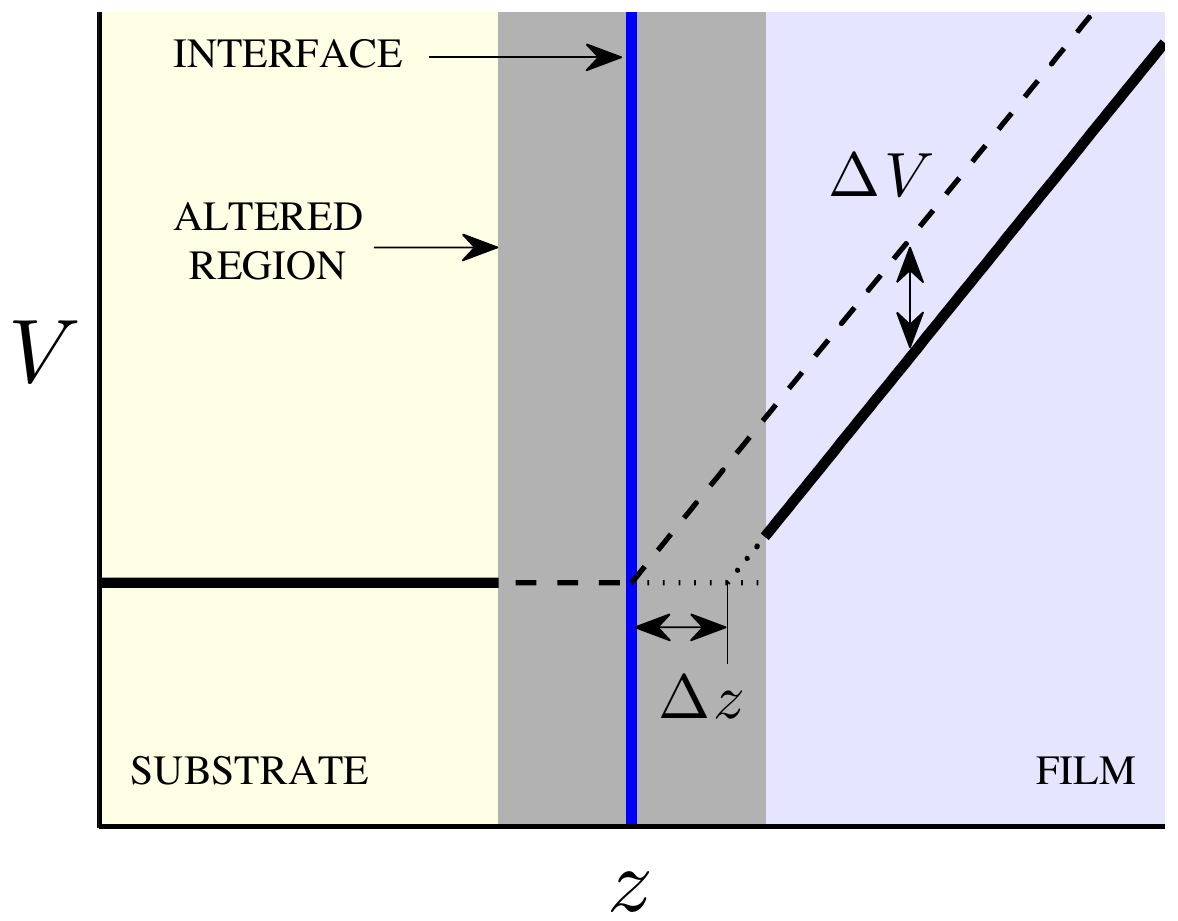}
\caption{\label{Interf1} Electrostatics near an 
interface with $\sigma_c$ net charge per unit area,
around which there is a region with cation intermixing
(dark grey band).
  Electrostatic potential energy $V$ 
versus distance perpendicular to the interface $z$.
The boundary condition is taken as constant potential
on the side of the substrate.
  The slope of the outgoing potential to the right is
$\sigma_c/\epsilon$ regardless of the inter-diffusion,
the latter producing a potential shift $\Delta V$, or 
apparent shift in interface position, $\Delta z$.}
\end{center}
\end{figure}

  The main effect of cation inter-diffusion across the 
interface is illustrated in Figure~\ref{Interf1}: the 
electrostatic potential is shifted with respect to the 
one arising from a pristine interface.
  It is essentially the effect of the interface dipole
generated by the swap of charges with its characteristic
potential drop $\Delta V$ (or raise).
  Alternatively, this effect can be interpreted as an effective
shift of the interface position, $\Delta z$ in the figure.
 This can be a productive way of seeing it, renormalising the
film thickness in the analysis in sections 1-4.
       
\begin{figure}
\begin{center}
\includegraphics[width=0.60\textwidth]{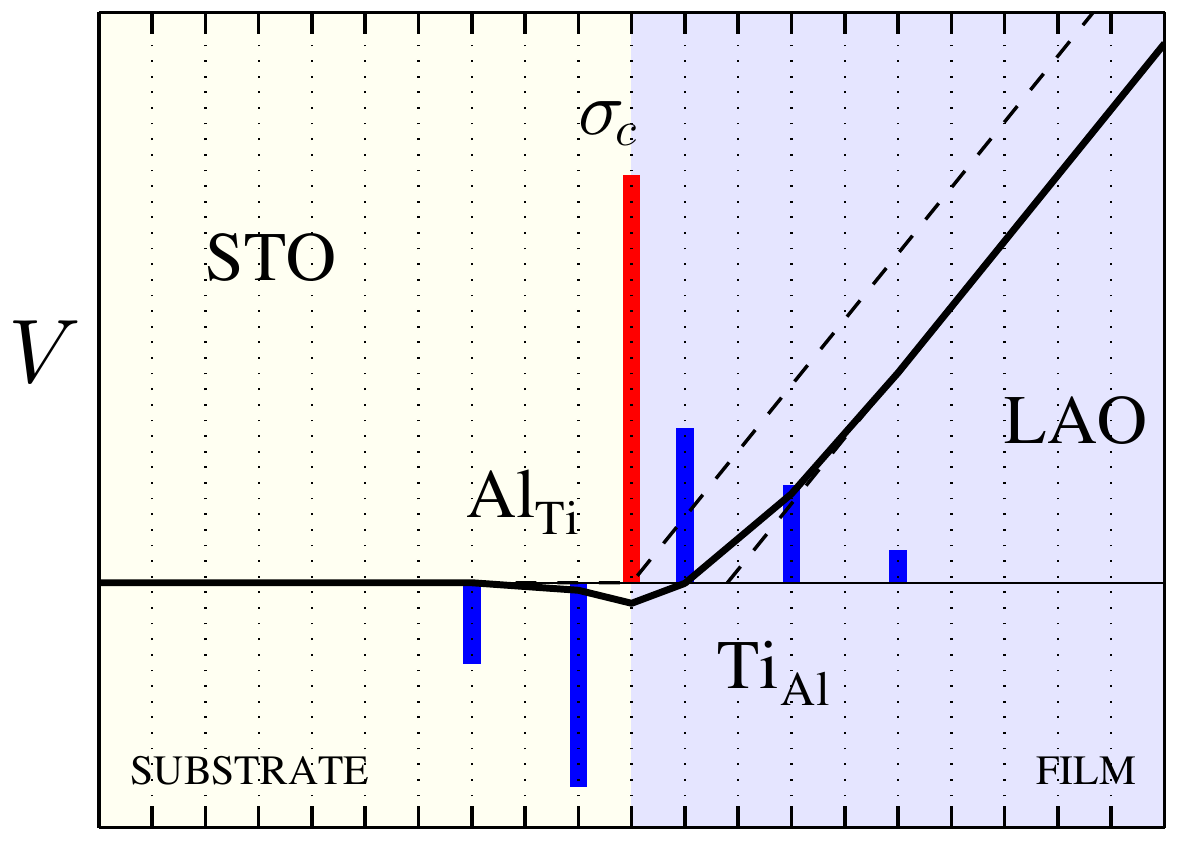}
\includegraphics[width=0.60\textwidth]{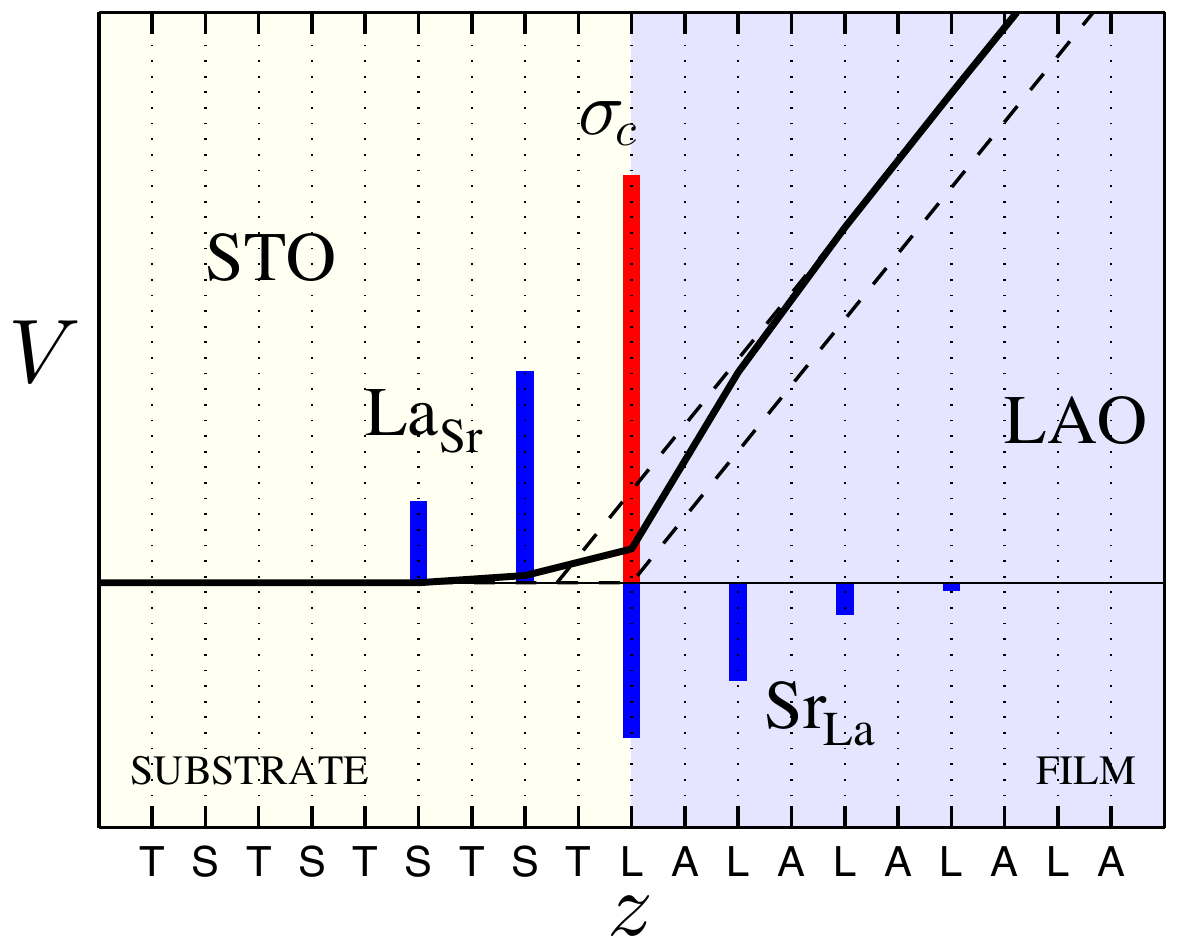}
\caption{\label{Interf2}  Electrostatic potential energy
for electrons, $V$, versus distance perpendicular to 
the (001) interface, $z$, for an interface with $\sigma_c$
net charge, and a distribution of charges in the
corresponding layers due to $B$ cation swapping
(upper panel) and $A$ cation swapping (lower panel).
T, S, L and A stand for the TiO$_2$, SrO, LaO and AlO$_2$
(001) planes respectively. 
$V$ is taken as constant on the side of the substrate.
 The (blue) bars indicate $\sigma_i$ values, the net 
charge in each layer (per formula unit area, $\Theta$) 
resulting from inter-diffusion.
  The (red) bar indicated by $\sigma_c$ sets the 
scale for the others.
  The potential energy changes by 1 eV in 
the whole range displayed on the vertical axis.
$\sigma_c = 0.5$ $e/\Theta$, 
$\epsilon_{\rm LAO}= 28$ and 
$\epsilon_{\rm STO}= 100$ 
(the latter chosen much smaller than actual values 
of about 300 for room temperature STO, in order to enhance the
$V$ variation on the STO side). 
}
\end{center}
\end{figure}

\subsubsection{Sign of the dipole and shift}

  The third point in this section is on the sign of
such effect. 
  Figure~\ref{Interf2} shows quantitative examples for
the behaviour of the electrostatic potential energy for
electrons in two cases of inter-diffusion, one for
$B$ cation exchange the other for $A$.
  The former, having Al and Ti ions exchanged, 
produces a voltage drop, or a right shift
of the effective interface position, and would thus
renormalise the film thickness to a smaller value.
  The latter produces the opposite. 
  From a thermodynamic point of view, $B$ cation 
inter-diffusion will be favoured by the electrostatics,
whilst $A$ cation inter-diffusion will tend to be 
suppressed
\footnote{Remember these arguments are based on the
original assumption of zero pristine interface dipole.
A large non-zero pristine dipole of appropriate sign
can reverse this trend.}.

\subsubsection{Scale of the interface shift}

  The fourth point is on the scale of interface shifts. 
  It is obvious that they will only be noticeable if on 
the scale of $\sigma_c$, i.e. if inter-diffusion is sizeable 
compared with 50\% per cation layer 
(inter-diffusion in the ppm scale can be happily ignored).
  Let us obtain a quantitative estimate for $\Delta z$.
  
  Let us consider the $(x,y)$-averaged charge density 
associated to inter-diffusion as $\delta_L(z)$ and
$\delta_R(z)$ for the charge at the left and the right sides
of the interface, respectively.
 The former with support
for $z<0$ only, the latter for $z \ge 0$ (we consider the
pristine interface at $z=0$).
  The fact that there are no external charges 
contributing to these densities is reflected in
\begin{equation}
\label{sigmaddef}
\sigma_d = - \int_{-\infty}^0 \! \! \delta_L(z) \, {\mathrm d}z = 
\int_0^{\infty} \! \! \delta_R(z) \, {\mathrm d}z   \; .
\end{equation}

  This description allows for $A$ or $B$ cation intermixing, 
and for both.
  $\sigma_d > 0$ implies more $B$-cation exchanges, as
in the upper panel of Figure~\ref{Interf2}, while $A$-cation
exchanges dominate for negative $\sigma_d$.

  The electric displacement at any $z$ position can be 
written as
\begin{equation}
D(z) = D(-\infty) + \int_{-\infty}^z \! \! \! 
\delta(z') \, {\mathrm d}z' \; .
\end{equation}
  The zero-field substrate boundary condition gives 
$D(-\infty)=0$.
  Assuming linear dielectric response, the electric field 
anywhere is given by $D(z) \equiv \epsilon_0 
{\cal E}(z) + P(z) = \epsilon(z)\, {\cal E}(z)$, and thus,
the field for any negative $z$ is given by
\begin{equation}
{\cal E}(z) = {1 \over \epsilon_{\rm STO} } 
\int_{-\infty}^z  \! \! \!  \delta_L(z') \, {\mathrm d}z '\; .
\end{equation}

  The electrostatic potential for anywhere in the 
substrate ($z<0$) is 
\begin{equation}
{\cal V}(z) = {\cal V}(-\infty) - \int_{-\infty}^z \! \! \! 
{\cal E}(z') \, {\mathrm d}z' \; .
\end{equation}
  Defining the reference potential as ${\cal V}(-\infty)=0$,
and using the previous expression for the field,
we  get (still for $z<0$)
\begin{equation}
{\cal V}(z) =  - {1\over \epsilon_{\rm STO} } \int_{-\infty}^z \! \! \!
 {\mathrm d}z'  \int_{-\infty}^{z'} \delta(z'') \, {\mathrm d}z''  \; ,
\end{equation}
which, reordering the double integral gives
\begin{eqnarray}
{\cal V}(z)  &=  - {1\over \epsilon_{\rm STO} } \int_{-\infty}^z \! 
\delta(z'') \, {\mathrm d}z''   \int_{z''}^z {\mathrm d}z'  \\
 &=  - {1\over \epsilon_{\rm STO} } \int_{-\infty}^z \! 
(z - z'') \delta(z'') \, {\mathrm d}z''   \\
 &= - {1\over \epsilon_{\rm STO} } \sigma_z 
\left ( z - \langle z \rangle_z \right ) \; , 
\end{eqnarray}
where $\sigma_z$ is the total charge integrated up to $z$,
and $\langle z \rangle_z$ is the average charge position
for the charge distribution up to $z$.
   The total potential change within the substrate up to
the interface is thus 
\begin{equation}
{\cal V}(0^-) = - \sigma_d { \langle z\rangle_{\rm STO} 
\over \epsilon_{\rm STO} }  \; , 
\end{equation}
$\langle z\rangle_{\rm STO}$ indicating the average position
of the diffusion charge in STO, and $\sigma_d$ as 
defined in Eq.~\ref{sigmaddef} (taking care of the sign).

  Within the LAO side the potential has two additive
components, namely, as originated by $\sigma_c$ and 
$\sigma_d$.
  The first one gives the straight line starting at $z=0$
in Figure~\ref{Interf2}, the second one saturates to 
a constant once $\delta_R(z)$ dies out (the sum of
both giving the potential curves in the figure).
  The shift of potential due to $\sigma_d$ in LAO
is found analogously to STO, giving the total potential
saturating at
\begin{equation}
{\cal V}(\infty) = \sigma_d \left (
{ \langle z\rangle_{\rm STO} \over \epsilon_{\rm STO} } -
{ \langle z\rangle_{\rm LAO} \over \epsilon_{\rm LAO} } 
 \right ) \; . 
\end{equation}
  The change in electrostatic potential energy for 
the electrons, as presented in Figure~\ref{Interf2},
$V=-e{\cal V}$, is thus
\begin{equation}
\Delta V = \sigma_d \, e \left (
{ \langle z\rangle_{\rm LAO} \over \epsilon_{\rm LAO} } -
{ \langle z\rangle_{\rm STO} \over \epsilon_{\rm STO} } 
 \right ) \; . 
 \end{equation}
  This expression reflects the dipole originated by 
the exchange, allowing a very intuitive interpretation of
such effect.
  Following the definition of the sign of $\sigma_d$,
$A$-cation exchange gives positive $\Delta V$, which is
negative for $B$-cation diffusion.

  As illustrated in Fig.~\ref{Interf1}, this effect can be 
described as a shift in the apparent position of the interface, 
$\Delta z$.
 Given the slope of the right-outgoing 
potential energy, $e \sigma_c / \epsilon_{\rm LAO}$, this gives
\begin{equation}
\label{last expression}
\Delta z = {\sigma_d \over \sigma_c}  \left ( 
\langle z \rangle_{\rm LAO}   - 
\langle z \rangle_{\rm STO}  {\epsilon_{\rm LAO} 
\over \epsilon_{\rm STO} } \right )  \; .
\end{equation}
  The thickness dependence of film properties, including
film critical thickness, should then refer to the effective
interface position, resulting in a net shift in the thickness 
axis of figures like Fig.~\ref{PHASES}.

  One consequence of Eq.~\ref{last expression} is that in a system
like LAO/STO, with $\epsilon_{\rm STO} \gg \epsilon_{\rm LAO}$, 
the width of the cation distribution in STO hardly affects the 
final shift
\footnote{Remember this prediction is under the original
assumption that $\epsilon$ is unaltered by intermixing 
either side of the film
which is likely inaccurate~\cite{Reinle-Schmitt2012}.
Deviations from such assumptions can be trivially added to the model.}
,
\begin{equation}
\Delta z \approx {\sigma_d \over \sigma_c} 
\langle z \rangle_{\rm LAO}  \; .
\end{equation}

  Considering now a LAO film of thickness $d$, this expression
implies that the field across the film will be noticeably screened
by cation inter-diffusion if ($i$) $\sigma_d \sim \sigma_c = +0.5$ 
$e/\Theta$, i.e. a very substantial amount of Ti$^{4+}$ going 
into LAO,  and ($ii$)  $\Delta z \sim d$, i.e. Ti cations penetrate 
the whole film. 
  Indeed, perfect screening would only be possible if half a 
monolayer of Ti cations all traversed the whole film to sit
on the surface, while their Al counterparts would go to 
the STO side staying close to the interface.
  Such cases were confirmed to screen the field within the calculations
of refs.~\cite{Chambers2010,Qiao2010,Yu2014}, and were obviously found
to be quite stable and even exothermic. 
Whether these swaps across the whole film actually happens depends 
on the kinetics of the process.






\subsection{Interfacial stoichiometry deviations \label{stoichiometry}}


  If in the growth process there is a cationic imbalance giving
rise to a net deviation of the compositional charge, the effect is
a trivial renormalisation of $\sigma_c$.
  Unlike the previous cases, where there were charge 
redistributions around the interface, with no change in the 
net charge, this section considers effects that alter the
total interfacial charge. 



 Recently several studies have discussed the role of 
La:Al off-stoichiometries, which can be modulated by the laser parameters of
the pulsed laser deposition,
on the carrier density
~\cite{Schoofs2011,Sato2013,Dildar2013,Breckenfeld2013,Warusawithana2013a}.
  It was shown that when the La/Al ratio differs from unity by just a few percent,
the carrier density can drop by two orders of magnitude. 
  It is clear that cation vacancies will affect $\sigma_c$, induce acceptor
 states and hence reduce the interface carrier density. 
  A concentration of just 2\% $A$-site vacancies across the
whole of a LAO film of 10 unit cells would correspond to a 
charge close to the half a quantum 0.5 $e/\Theta$. 
 Further discussion is beyond the scope of this paper and we refer 
the reader to more detailed studies~\cite{Seo2011,Gu2012} 
of various types of non-stoichiometric LAO-STO systems.
  
  These studies show that several
types of defect structures are thermodynamically stable,
for example the combined La vacancy at the LaO interface
and O vacancy at the AlO$_2$ surface, which
screen the internal electric field and in some cases 
remove the interface carriers. 
  In fact, this last kind of combined defects at surface and
interface represent a similar charge screening mechanism
to the ones modelled within this review: instead of the scenario of surface O 
vacancies and interface free electron carriers, it is surface
O vacancies and interface cation vacancies. 
  The equilibrium occurrence of such combined defects
can be treated with the same model (the $C$ parameter
would now include the combined formation energy of both 
defects), but the key difference is that this mechanism does
not produce carriers for the 2DEG.

  It is likely that in several systems there also exists a total 
deficit of oxygen, via the creation of oxygen vacancies
in the substrate during the growth process or by other means,
even if they are not stabilised or favoured by the electrostatics.
  Here each vacancy, which is likely created near the 
substrate surface, creates two electrons which are 
loosely bound near the vacancy.
  These processes have been suggested to play a 
role in LAO-STO when grown under insufficient oxygen 
partial pressure~\cite{Basletic2008}.
 They may also be created at the bare 
STO surface when subject to ultraviolet light 
irradiation~\cite{Meevasana2011} or vacuum 
cleaved~\cite{Santander-Syro2011}.

  The unexpected conductivity observed at the non-polar 
interface between an \textit{amorphous} LAO film and a 
crystalline STO substrate~\cite{Herranz2012,Chen2011,
Lee2012,Christensen2013,Liu2013} has recently also been 
explained as the creation of oxygen vacancies in the STO 
substrate during the growth process~\cite{Lee2012,Liu2013}.
  If the system is subjected to post-growth annealing in oxygen,
which presumably removes these vacancies, the conductivity 
is no longer observed~\cite{Liu2013}. 
  These processes are also possibly happening in other 
heteroepitaxial non-polar oxide systems~\cite{Chen2013}.

Interestingly alterations in the net charge
can also arise if there are half-unit-cell steps at the interface,
with both $n$ and $p$ interface terraces.
  In such case there is a continuous variation of $\sigma_c$, 
from $-0.5 e/\Theta$ to $+0.5 e/\Theta$, depending on the ratio of 
interface area covered by each kind, including a total
removal of the polarisation discontinuity for the case
of the interface being composed by half $n$ and half $p$
interfaces. 
  For the cancellation to be effective, though, the 
width of the terraces should be smaller than the film
thickness.



 \section{Summary}

  In this paper we have reviewed the present understanding
on the origin of a 2DEG observed at the epitaxial interface between
certain perovskite insulator thin films and certain perovskite
insulator substrates, referring mostly to the paradigmatic case
of an LAO film on STO.
  The ``polar catastrophe" concept and existing evidence for and
against it have been reviewed.
  The polarisation discontinuity of half a quantum of polarisation 
(for LAO on STO) is obtained from the surface theorem of
Ref.~\cite{Vanderbilt1993}, and the symmetry of the bulk systems.
  This construction is related to other symmetry-protected topological 
phases appearing in other contexts, namely, the topological insulators
related to other symmetries as time reversal and/or particle-hole
symmetry.
  The discussion is generalised from the conventional (001)
orientation of the interface to any orientation via the lattice
of meaningful polarisation vectors for these insulators, 
including high-index vicinal (stepped) interfaces.
  Alternative descriptions and formulations for the polar discontinuity
found in the literature are reviewed and compared with the
one based on formal polarisation.

  The polar catastrophe makes the pristine film unstable 
towards electric-field screening mechanisms based on 
surface and interface charges, the latter being the
electron carriers of the 2DEG.
  The review of such instabilities is upheld by a simple
model, following on Refs.~\cite{Bristowe2009,
Bristowe2011}, starting from the electronic reconstruction,
followed by the redox-defect model, and their possible
coexistence. 
  The model is generalised to arbitrary homogeneous 
surface-interface charging mechanisms. 

  Finally, the effect of deviations from the ideal epitaxial
growth (interface effects) are reviewed, distinguishing between 
deviations respecting stoichiometry (such as inter-diffusion across the
interface), which do not alter the qualitative situation described
by the polar catastrophe and surface-interface charge screening,
and deviations that do alter stoichiometry on a substantial scale.

\ack
  We wish to thank H. Y. Hwang, S. A. T. Redfern, P. Ordej{\'o}n, P. Aguado,
I. Souza, R. Resta, D. Fontaine and D. Vanderbilt for useful discussions.
We acknowledge support of EPSRC and computing 
resources of CAMGRID in Cambridge and the 
Spanish Supercomputer Network (RES).
  This work has been partly funded by UK's EPSRC 
and the ARC project TheMoTherm (Grant No. 10/15-03).
  Work at Argonne supported by DOE-DES under Contract 
No. DE-AC02-06CH11357.
  PhG acknowledges a Research Professorship of the 
Francqui Foundation (Belgium).

\section*{References}


\providecommand{\newblock}{}

\end{document}